\pdfoutput=1

\documentclass[12pt]{report}
\usepackage{graphics}
\usepackage[round,authoryear]{natbib}
\usepackage{rotating}
\usepackage{tocloft}
\usepackage{color}

\newcommand{\singlespace}{\setlength{\baselineskip}{0.165in}}
\newcommand{\doublespace}{\setlength{\baselineskip}{0.33in}}

\newcommand{\CHAPTER}[1]{%
\refstepcounter{chapter}%
\addcontentsline{toc}{chapter}{\thechapter . #1}%
\newpage%
\pagestyle{myheadings}
\thispagestyle{plain}%
\singlespace%
\begin{center}%
\LARGE
\textbf{\thechapter}%
\textbf{. #1}\\%
\normalsize
\end{center}%
\doublespace%
\vspace{2em}
\singlespace}

\newcommand{\TOPPER}[1]{%
\addcontentsline{toc}{chapter}{#1}%
\newpage%
\pagestyle{plain}%
\begin{center}%
\LARGE
\textbf{#1}\\%
\normalsize
\end{center}%
\vspace{2em}
\singlespace}

\newcommand{\gtrsim}{\mathrel{\hbox{\rlap{\hbox{\lower4pt\hbox{$\sim$}}}\hbox{$>$}}}}
\newcommand{\lesssim}{\mathrel{\hbox{\rlap{\hbox{\lower4pt\hbox{$\sim$}}}\hbox{$<$}}}}

\begin{document}
\pagenumbering{roman}
\thispagestyle{empty}
\begin{center}
REDUCING FALSE ALARMS IN SEARCHES FOR GRAVITATIONAL WAVES  FROM COALESCING BINARY SYSTEMS

\vspace{1.75in}

A Thesis\\
\vspace{1ex}
Submitted to the Graduate Faculty of the\\
Louisiana State University and\\
Agricultural and Mechanical College\\
in partial fulfillment of the\\
requirements for the degree of\\
Master of Science\\
\vspace{1ex}
in\\
\vspace{1ex}
The Department of Physics and Astronomy\\

\vspace{1.75in}

by\\
Andr\'es Rodr\'iguez\\
B.S., Florida State University, 2003\\
August, 2007\\

\end{center}

\singlespace
\TOPPER{Dedication}
\doublespace
I dedicate this thesis to all the loved ones who have passed away during the course of my life: Abuelo, Miriam, Tio Fran, Nina, Alfredo, and Elena.

\singlespace
\TOPPER{Acknowledgments}
\doublespace
I have been very fortunate to have had the opportunity to collaborate with and learn from so many amazing scientists over the past several years. I particularly want to thank my advisor Professor Gabriela Gonz\'alez  - a great mentor and a true inspiration to how science should be done, to Professor Jorge Pullin - I will never forget the email I received from you notifying me that I was accepted to LSU and the opportunity for the Sloan Foundation fellowship, to Chad Hanna - thanks for letting me drop by your office next door countless times concerning C and MATLAB, for collaborating on the PBH searches, and your editorial help with the thesis. I'd like to also thank the LSU Experimental General Relativity group, particularly Rupal Amin, Jacob Slutsky, Dr.\  Ravi Kopparapu, Professor Warren Johnson, and Andrew Weber. From LIGO, I'd like to thank Professor Peter Shawhan - thanks for the initial inspiration of what would become the $r^2$ test,  the Compact Binary Coalescence (CBC) Group, particularly Dr.\ Duncan Brown, Dr.\ Stephen Fairhurst, Dr.\ Eirini Messaritaki, Professor Patrick Brady, Dr.\ Alexander Dietz, and Drew Keppel for all of their help with the inspiral analysis, help with understanding the wonderful world of LAL/LALapps, and various discussions on LIGO data analysis. I would like to thank both the LIGO Scientific Collaboration (LSC) and the CBC group as a whole for access to LIGO data. I would also like to thank the LSC inspiral review committee for the review of the S3 PBH search. 

From my years at Florida State University I'd like to thank the scientists and engineers I had the privilege of collaborating with including Professor Vasken Hagopian, Professor James Skofronick, Professor Sanford Safron, Dr.\ Elshan Akhadov, and Maurizio Bertoldi. 

I thank the Louisiana State University Department of Physics \& Astronomy for the opportunity to attend graduate school here and the Sloan Foundation for the fellowship award I received to help fund my studies during the years here at LSU. 

Finally, I'd like to thank my wonderful family including my mom Cecilia, my dad Andy, my sister Aileen, my brother Alex, and my abuela Aida for all their love and support all my life. I'd also like to thank my beautiful girlfriend Kim for all her love and support during this past year. 

This research has been supported by National Science Foundation grants PHY-0135389, PHY-0355289.

\singlespace
\TOPPER{Table of Contents}
\singlespace
\contentsline {chapter} {Dedication}{ii}
\doublespace
\singlespace
\contentsline {chapter} {Acknowledgments}{iii}
\doublespace

\singlespace
\contentsline {chapter}{List of Tables}{vii}
\doublespace

\singlespace
\contentsline {chapter}{List of Figures}{viii}
\doublespace

\singlespace
\contentsline {chapter}{Abstract}{x}
\doublespace

\singlespace
\contentsline {chapter}{1. Introduction}{1}
\doublespace

\singlespace
\contentsline {chapter}{2. Gravitational Radiation}{3}
\contentsline {section}{\numberline {2.1}Gravitational Waves}{3}
\contentsline {section}{\numberline {2.2}Astrophysical Sources of Gravitational Waves}{5}
\contentsline {section}{\numberline {2.3}Laser Interferometer Gravitational Wave Observatory (LIGO)}{7}
\contentsline {subsection}{\numberline {2.3.1}Design}{8}
\contentsline {subsection}{\numberline {2.3.2}Noise Sources}{9}
\contentsline {subsection}{\numberline {2.3.3}Science Runs}{11}
\doublespace

\singlespace
\contentsline {chapter}{3. Data Analysis for Coalescing Binary Systems}{13}
\contentsline {section}{\numberline {3.1}Optimal Matched Filtering}{13}
\contentsline {section}{\numberline {3.2}Searching for Gravitational Waves from Coalescing Binary Systems: The Inspiral Pipeline}{17}
\contentsline {section}{\numberline {3.3}Tuning the Search for Inspiralling Binaries}{19}
\contentsline {section}{\numberline {3.4}Vetoes}{22}
\contentsline {subsection}{\numberline {3.4.1}Instrumental Vetoes}{23}
\contentsline {subsection}{\numberline {3.4.2}Statistical Vetoes}{24}
\contentsline {subsection}{\numberline {3.4.3}Signal Based Vetoes}{24}

\doublespace

\singlespace
\contentsline {chapter}{4. Methods to Reduce False Alarms in Coalescing Binary System Searches}{25}
\contentsline {section}{\numberline {4.1}The $\chi^2$ Veto}{25}
\contentsline {section}{\numberline {4.2}A Test to Further Reduce False Alarms}{27}
\contentsline {section}{\numberline {4.3}Implementation of the r$^2$ Test Into the Inspiral Pipeline}{34}
\doublespace

\singlespace
\contentsline {chapter}{5.The Search for Primordial Black Holes in LIGO's 3rd Science Run}{36}
\contentsline {section}{\numberline {5.1}The Third LIGO Science Run}{36}
\contentsline {section}{\numberline {5.2}Tuning the Search}{39}
\contentsline {section}{\numberline {5.3}Results for the r$^2$ Test}{43}
\contentsline {section}{\numberline {5.4}S3 PBH Search Result}{46}

\doublespace

\singlespace
\contentsline {chapter}{6. The r$^2$ Test in Other LIGO Searches}{48}
\contentsline {section}{\numberline {6.1}S3$/$S4 Binary Neutron Star Searches}{48}
\contentsline {section}{\numberline {6.2}S4 Primordial Black Hole Search}{51}
\contentsline {section}{\numberline {6.3}S5 Binary Black Hole Search (Epoch 1)}{53}

\doublespace

\singlespace
\contentsline {chapter}{7. Conclusions}{56}
\doublespace

\singlespace
\contentsline {chapter}{Bibliography}{57}
\doublespace
\singlespace
\contentsline {chapter}{Vita}{60}

\clearpage
\singlespace
\TOPPER{List of Tables}
\doublespace
\contentsline {table}{\numberline {5.1}{\ignorespaces Times Analyzed When at Least Two Detectors Were Operating}}{39}
\contentsline {table}{\numberline {5.2}{\ignorespaces Summary of the S3 PBH Coincidence Windows}}{41}
\contentsline {table}{\numberline {5.3}{\ignorespaces r$^2$ Test Parameters for the S3 PBH Search}}{45}
\contentsline {table}{\numberline {6.1}{\ignorespaces Target Sources of the S3/S4 BNS Search}}{48}
\contentsline {table}{\numberline {6.2}{\ignorespaces Summary of the S3/S4 BNS Coincidence Windows}}{49}
\contentsline {table}{\numberline {6.3}{\ignorespaces r$^2$ Test Parameters for the S3/S4 BNS Search}}{49}
\contentsline {table}{\numberline {6.4}{\ignorespaces  r$^2$ Test Results for S3/S4 BNS Searches}}{50}
\contentsline {table}{\numberline {6.5}{\ignorespaces Target Sources of the S4 PBH Search}}{51}
\contentsline {table}{\numberline {6.6}{\ignorespaces  Summary of the S4 PBH Coincidence Windows}}{51}
\contentsline {table}{\numberline {6.7}{\ignorespaces  r$^2$ Test Parameters for the S4 PBH Search}}{51}
\contentsline {table}{\numberline {6.8}{\ignorespaces  r$^2$ Test Results for S4 PBH Search}}{52}
\contentsline {table}{\numberline {6.9}{\ignorespaces Target Sources of the S5 BBH Search}}{53}
\contentsline {table}{\numberline {6.10}{\ignorespaces Summary of the S5 BBH Coincidence Windows}}{53}
\contentsline {table}{\numberline {6.11}{\ignorespaces r$^2$ Test Parameters for the S5 BBH Search}}{53}
\contentsline {table}{\numberline {6.12}{\ignorespaces r$^2$ Test Results for the S5 BBH Epoch 1}}{54}
\vspace{1em}

\clearpage
\singlespace
\TOPPER{List of Figures}
\doublespace
\contentsline {figure}{\numberline {2.1}{\ignorespaces Effect of a Gravitational Wave On a Ring of Particles}}{5}
\contentsline {figure}{\numberline {2.2}{\ignorespaces The Two LIGO sites}}{7}
\contentsline {figure}{\numberline {2.3}{\ignorespaces Simplified Schematic of LIGO}}{8}
\contentsline {figure}{\numberline {2.4}{\ignorespaces Noise Limits for the LIGO Design}}{10}
\contentsline {figure}{\numberline {2.5}{\ignorespaces Best Strain Sensitivities for the LIGO 4km Interferometers}}{12}
\contentsline {figure}{\numberline {3.1}{\ignorespaces Example BNS Waveform}}{15}
\contentsline {figure}{\numberline {3.2}{\ignorespaces Illustration of the Inspiral Coincidence Analysis Pipeline}}{20}
\contentsline {figure}{\numberline {3.3}{\ignorespaces Illustration of How Single Interferometer Data is Divided}}{21}
\contentsline {figure}{\numberline {4.1}{\ignorespaces Visual Representation of $\chi^2$ Statistic}}{26}
\contentsline {figure}{\numberline {4.2}{\ignorespaces Simulated Inspiral Injection SNR and Weighted $\chi^2$ Time Series}}{28}
\contentsline {figure}{\numberline {4.3}{\ignorespaces SNR and Weighted $\chi^2$ Time Series for a Segment of Data}}{29}
\contentsline {figure}{\numberline {4.4}{\ignorespaces Histogram of Figure 4.3 SNR and Weighted $\chi^2$ Time Series}}{30}
\contentsline {figure}{\numberline {4.5}{\ignorespaces Comparison of SNR and Weighted $\chi^2$ Time Series for False Alarm and Injection}}{31}
\contentsline {figure}{\numberline {4.6}{\ignorespaces Illustration of the r$^2$ Test}}{32}
\contentsline {figure}{\numberline {4.7}{\ignorespaces r$^2$ Test Example Result}}{33}
\contentsline {figure}{\numberline {5.1}{\ignorespaces LIGO S3 Run Best Strain Sensitivities for All Interferometers}}{37}
\contentsline {figure}{\numberline {5.2}{\ignorespaces Example PBH Waveform}}{38}
\contentsline {figure}{\numberline {5.3}{\ignorespaces SNR of Coincident Time Slides and Coincident Injections}}{40}
\contentsline {figure}{\numberline {5.4}{\ignorespaces Effective SNR Statistic}}{41}
\contentsline {figure}{\numberline {5.5}{\ignorespaces Timing Coincidence ($\Delta$T) Histogram}}{42}
\contentsline {figure}{\numberline {5.6}{\ignorespaces $\Delta\mathcal{M}_c$ Histogram}}{42}
\contentsline {figure}{\numberline {5.7}{\ignorespaces $\Delta$$\eta$ Histogram}}{43}
\contentsline {figure}{\numberline {5.8}{\ignorespaces H1H2 Effective Distance Cut}}{44}
\contentsline {figure}{\numberline {5.9}{\ignorespaces Results of the r$^2$ Test  for the S3 PBH Search}}{45}
\contentsline {figure}{\numberline {5.10}{\ignorespaces Foreground Versus Background Distribution of Triggers}}{47}
\contentsline {figure}{\numberline {6.1}{\ignorespaces Results of the r$^2$ Test for the S3 BNS Search}}{49}
\contentsline {figure}{\numberline {6.2}{\ignorespaces Results of the r$^2$ Test for the S4 BNS Search}}{50}
\contentsline {figure}{\numberline {6.3}{\ignorespaces Results of the r$^2$ Test for the S4 PBH Search}}{52}
\contentsline {figure}{\numberline {6.4}{\ignorespaces Example BBH Waveform}}{54}
\contentsline {figure}{\numberline {6.5}{\ignorespaces Results of the r$^2$ Test for the S5 BBH Search}}{55}
\vspace{1em}

\clearpage

\singlespace
\TOPPER{Abstract}
\doublespace
LIGO observatories in Livingston, LA and Hanford, WA may detect gravitational waves emitted from coalescing binary systems composed of two compact objects. In order to detect compact binary coalescence (CBC) events, LIGO searches utilize matched filtering techniques. Matched filtering is the optimal detection strategy for stationary, Gaussian noise, however, LIGO noise is often non-stationary, non-Gaussian. Non-stationary noise result in an excess of false candidate events, commonly known as false alarms. This thesis develops the r$^2$ test to reduce the false alarm rate for LIGO CBC searches. Results of the search for primordial black hole binary systems (where each object has less than 1M$_\odot$), in LIGO's Third Science Run (S3) is also presented.

Results of the r$^2$ test are shown for several LIGO CBC searches, including the binary neutron star searches in the Third and Fourth Science Runs (S3$/$S4), the S3$/$S4 primordial black hole searches, and the binary black hole search in the first three months of the Fifth Science Run (S5). The r$^2$ test significantly reduces the false alarm rate in these searches, while only falsely dismissing a small fraction of simulated events.

\pagenumbering{arabic}
\clubpenalty=10000
\widowpenalty=10000

\pagestyle{myheadings}
\CHAPTER{Introduction}
\label{chapter1}
\doublespace
At present time, a worldwide network of interferometric detectors are searching for gravitational waves. These include LIGO \citep{initialLIGO} - which consists of two observatories with three detectors located in Hanford, WA USA (2km, 4km) and Livingston, LA (4km) USA; VIRGO \citep{VIRGO} - a 4km interferometer located in Pisa, Italy; and GEO \citep{GEO} - a 600m interferometer located in Hannover, Germany.  

This thesis contributes to the search of gravitational waves, specifically those emitted from coalescing binary systems composed of two compact objects (such as a neutron stars or black holes), by providing a new method the reduce the rate of false candidate events. The method is called the r$^2$ test and will be introduced in chapter four. Results of the search for primordial black hole binary systems (where each object has less than 1M$_\odot$), in LIGO's Third Science Run (S3) is also presented in chapter five. This thesis specifically uses data from the LIGO interferometers, although methods described could be used with data from the other gravitational wave detectors mentioned. 

Chapter two introduces gravitational radiation, gravitational waves, astrophysical sources of gravitational waves, and the Laser Interferometer Gravitational Wave Observatory (LIGO). 

Chapter three describes data analysis for coalescing binary systems, including the theory of matched filtering, methods for detecting gravitational waves from coalescing binary systems, tuning searches for coalescing binary systems, and vetoes.

In the fourth chapter, we describe methods to reduce false alarms in coalescing binary system searches, including the $\chi^2$ veto, and a test to further reduce false candidate events, known as the r$^2$ test. The last section describes how the r$^2$ test is implemented into the current searches of gravitational waves using LIGO data.

The fifth chapter summarizes the search for primordial black holes in the Third LIGO Science Run (S3), including the tuning of the search, results for the r$^2$ test, and the result of the PBH binaries search in S3.

The sixth chapter summarizes results for the r$^2$ test in several other LIGO searches including the S3/S4 binary neutron star searches, S4 primordial black hole search, and the S5 binary black hole search (epoch 1). 

\emph{The analysis presented in this thesis is the fruit of my own work, in collaboration with the Compact Binary Coalescence (CBC) analysis group, which includes members of the LIGO Scientific Collaboration (LSC) and Virgo. The results are under review by the LSC, thus they are subject to possible revision before journal publication and they do not necessarily reflect the opinions of the LIGO Scientific Collaboration, the CBC group, or Louisiana State University.}

\CHAPTER{Gravitational Radiation}
\label{chapter2}
\doublespace

This chapter describes gravitational waves, their sources, and an experiment presently conducted to detect these waves. In section \ref{gravitywaves}, we will provide the basis of gravitational waves physics as described by Einstein's theory of General Relativity. Section \ref{gravitywavessources} discusses the various sources of gravitational radiation, and in section \ref{LIGO} we describe detectors being operated today to discover gravitational waves directly, the Laser Interferometer Gravitational Wave Observatory (LIGO).

\section{Gravitational Waves}
\label{gravitywaves}
   Einstein's theory of General Relativity \citep{Einstein} can be summed up in this simple, yet powerful equation:
\begin{equation}
{\rm G_{\mu\nu}}=8\pi G \rm{T}_{\mu\nu},
\end{equation}
known as Einstein's equation. It essentially tells us how the curvature of spacetime, coded in the Einstein tensor G$_{\mu\nu}$, reacts to the presence of matter/energy, coded in T$_{\mu\nu}$, the energy momentum tensor, where $G$ is Newton's constant \citep{Thorne87}. Since G$_{\mu\nu}$ and T$_{\mu\nu}$ are symmetric 4-tensors, this equation is actually a set of 10 coupled non-linear partial differential equations. Few exact solutions are analytically known due to the non-linearity of the equations. One way of gaining intuition about physical solutions to the field equations is to make approximations. For example, in the weak field limit we assume the gravitational field is very weak and the spacetime approximates that of Minkowski space. The metric, $g_{\mu\nu}$, can be written as:
 \begin{equation}
{g_{\mu\nu}} = \eta_{\mu\nu} + h_{\mu\nu}, \qquad  |h_{\mu\nu}| << 1 
\end{equation}
where $\eta$$_{\mu\nu}$ is the flat Minkowski metric and $h_{\mu\nu}$ is the small metric perturbation, which will influence the geometry of space-time, and thus the motion of particles and how light travels. The weak field limit itself transforms the non-linear Einstein equations in vacuum (T$_{\mu\nu}$ = 0) into linear equations, with further simplifications using the transverse traceless (TT) gauge, therefore making Einstein's equation become a wave equation:
\begin{equation}
\Bigg(\bigtriangledown^2 - \frac{1}{c^2}\frac{\partial^2}{\partial t^2}\Bigg)  h_{\mu\nu} = 0
\end{equation}

 For a a wave propagating in the \^ z direction, $h_{\mu\nu}$ is:
\begin{equation}
{h_{\mu\nu}} =
 \left(
\begin{array}{cccc} 	
0 & 0 & 0 & 0\\ 	
0 & h_{+} & h_{\times} & 0 \\ 	
0 & h_{\times} & -h_{+} & 0\\ 	
 0 & 0 & 0 & 0\\ 	
\end{array} 	
\right)
\end{equation}
where h$_{+}$(t) and h$_{\times}$(t) are the two independent polarizations 
of the gravitational wave. For gravitational waves being emitted from a binary composed of two neutron stars, the polarizations of the gravitational wave are given by \citep{Duncan_thesis}:
\begin{equation} 
\label{hplus}
h_{+}(t) = -\frac{2G}{c^{4}r}\mu(\pi GMf)^{2/3}(1 + \cos^{2}\iota)\cos(2\pi ft - 2 \phi_{0})
\end{equation} 
\begin{equation} 
\label{hcross}
h_{\times}(t)  = -\frac{4G}{c^{4}r}\mu (\pi GMf)^{2/3} \cos\iota \ \sin(2\pi ft - 2\phi_{0}) 
\end{equation} 
where $\iota$ is the inclination angle of the source, $\phi_{0}$ is the initial orbital phase, $f$ is the frequency of the gravitational wave, $r$ is the distance from the detector to the binary system, $M$ is the total mass of the binary system, $M = m_{1} + m_{2}$,  $\mu$ is the reduced mass of the system, $\mu$ = m$_1$m$_2$ $/$ (m$_1$+m$_2$). The frequency of the gravitational wave, $f$, evolves as the two neutron stars inspiral toward one another and is given by \citep{Duncan_thesis, rsqref}:
\begin{equation} 
f = \frac{5^{3/8}c^{15/8}}{8\pi G^{5/8}\mu^{3/8}M^{1/4}(t_{c}-t)^{3/8}}
\end{equation} 
where c is the speed of light and $t_c$ is the coalescence time. 

\begin{figure}[!hbp|t]
\begin{center}
   \centering
   \includegraphics[height=6cm,width=10cm,angle=0]{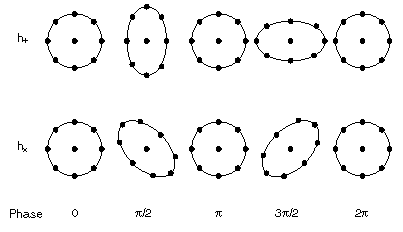} 
   \caption{The effect of a gravitational wave on a ring of particles. The wave propagation is perpendicular to the ring. The top row shows the effect of the wave $+$ polarized, while the bottom row shows the effect of the $\times$ polarization \citep{Merkowitz}.}
   \label{ring}
\end{center}
\end{figure}

  Physically speaking, what is the effect of this gravitational wave passing by an object? A classic example is to imagine a ring of particles in the x-y plane. As the gravitational wave propagates along the \^z axis, $h_{+}$ would stretch the distances between the particles along \^x axis while shrinking the distances in the \^y axis and vice-versa. The  $h_{\times}$ polarization of the wave would have the same effect, although it would be rotated by 45$^\circ$. This is shown in figure \ref{ring}.
  
  What type of astrophysical phenomena can cause a gravitational wave? This is described in the following section. 
  
\section{Astrophysical Sources of Gravitational Waves}
\label{gravitywavessources}

There are a large range of astrophysical phenomena that could produce gravitational waves strong enough plausibly detectable by ground based interferometers. An example of such detectors is the LIGO, where phenomena that produce gravitational waves within LIGO's sensitive band, the 40Hz-4kHz, could be detected in the near future if the signal is strong enough to be above noise and is distinguishable from false alarms. 
   
The first category of gravitational wave sources include those that emit brief, broadband signals, such as those produced from the collapse of a supernova or coalescence of binary systems, called \textit{burst signals}. In particular, Type II supernova collapses can yield strong gravitational waves (h$\sim$$10^{-20}$). Another example of a burst source source includes those producing gamma ray bursts (GRBs), where sources are speculated to be from a variety of astrophysical phenomena ranging from black hole mergers to solar flares. 

A second category includes \textit{periodic signals}, resulting from rotating stars, such as pulsars. Studies performed on the first discovered binary pulsar (PSR 1913+16) by Russell Hulse and Taylor in 1974 provide the best empirical evidence we have of gravitational waves indirectly to date \citep{HulseTaylor}. This was done by studying the orbital period of the binary system and accurately predicted the speeding up as predicted by general relativistic emission of gravitational waves.

The third source of gravitational waves is a \textit{stochastic background}, gravitational waves emitted from the beginnings of the early universe (cosmological stochastic background) or from a large an ensemble of unresolved astrophysical sources. 

Gravitational waves can also be emitted by \textit{compact binary star systems}, composed of neutron stars or black holes. General Relativity predicts that the shrinking of the binary orbit of these systems is due to emission of gravitational radiation, with the gravitational waves being emitted at twice the orbital frequency of the binary. This process continues until the orbit begins to rapidly shrink (inspiral) and gravitational waveform becomes a ``chirp'' signal, increasing in frequency and amplitude. To give a sense of the time scale for this process, imagine an inspiralling system composed of two 1.4 M$_{\odot}$ traversing through LIGO's sensitive band with a starting frequency of 40Hz, the signal would spend about 25 seconds in LIGO's sensitive band. In the case of low mass binary systems, we can approximate the waveform using the 2$^{nd}$ order Post-Newtonian expansion \citep{Blanchet, Damour}. We will return to the discussion of coalescing binary systems and methods used to detect these signals in chapter \ref{chapter3}. 

\renewcommand{\baselinestretch}{1}
\section{Laser Interferometer Gravitational Wave Observatory (LIGO)}
\label{LIGO}
\begin{figure}[!hbp|t]
\begin{center}
   \centering
   \includegraphics[height=8cm,width=12cm,angle=0]{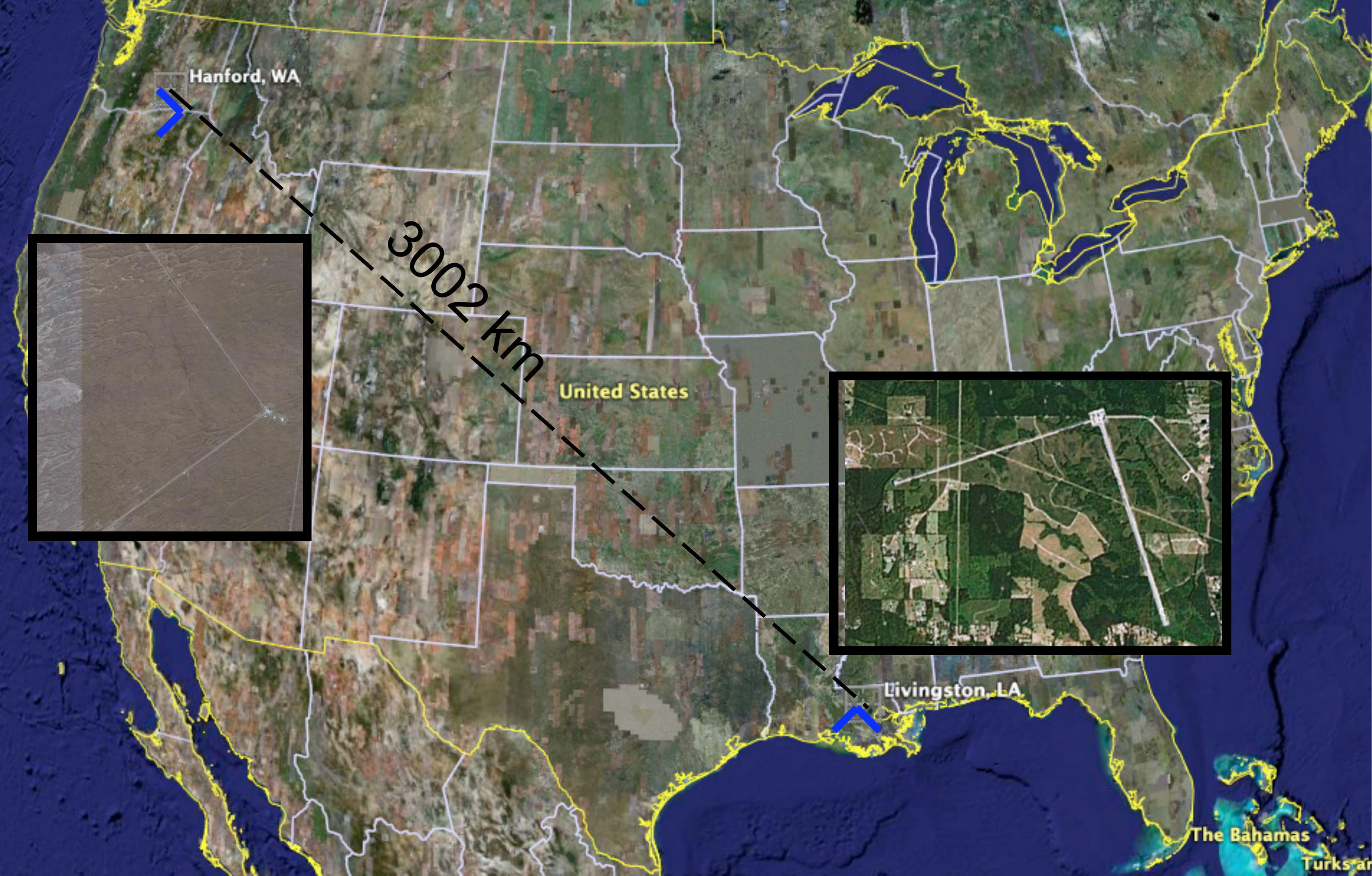} 
   \caption{The two LIGO sites \citep{Google}.}
   \label{LIGOsites}
\end{center}
\end{figure}

 \renewcommand{\baselinestretch}{2}
The Laser Interferometer Gravitational Wave Observatory (LIGO) is a joint Caltech and MIT project supported by the National Science Foundation. The observatory is composed of two sites, one in Hanford, WA and the other in Livingston, LA. The LIGO Hanford Observatory (LHO) contains both a 4km (H1) and 2km (H2) interferometers which are independent of one another, yet share the same vacuum system. The LIGO Livingston Observatory (LLO) contains one 4km interferometer (L1). Each of the interferometer arms are perpendicular to one another. The orientation of the Hanford Observatory is closely aligned to the Livingston Observatory, as to give a common response to a gravitational wave signal, the North-South arms of both detectors are along a great circle. This is depicted in figure \ref{LIGOsites}. 

Beyond the two LIGO sites, an international team of scientists has been created in order to analyze the data and to contribute to the improving of the experiment. This group is called the LIGO Scientific Collaboration (LSC) \citep{LSC}. It includes scientists working at the LIGO laboratory sites and scientists representing over 52 institutions from around the world, including Louisiana State University. 

\subsection{Design}
\label{design}

\begin{figure}[!hbp|t]
\begin{center}
   \centering
   \includegraphics[height=7cm,width=8cm,angle=0]{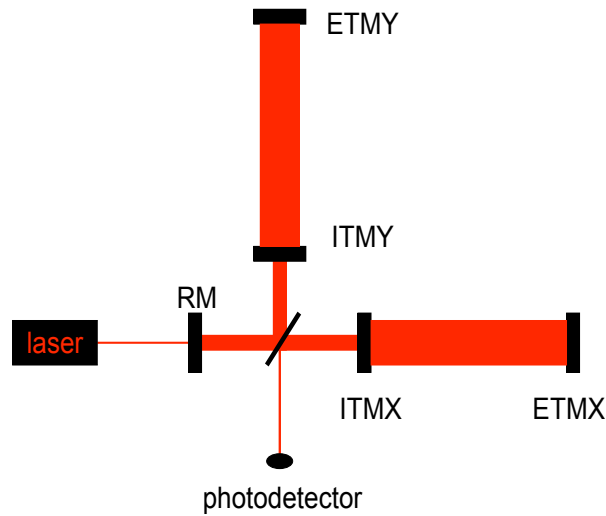} 
   \caption{\singlespace A simplified schematic of LIGO interferometer. RM is the recycling mirror, ITMX,Y are the intermediate test mass (partially transmitting mirrors), ETMX,Y are the end test mass (mirrors).}
   \label{LIGOsimple}
\end{center}
\end{figure}

The LIGO detector design is a power-recycled Michelson interferometer with Fabry-Perot arm cavities, which have a set of equal length perpendicular arms of 4 km. The laser light source (Nd:YAG laser, operating at a wavelength of 1.06 $\mu$m with a maximum output power of 10W) is sent to the beam splitter which directs the light along the two perpendicular arms, which first pass through a partially transmitting input mirror (ITM) which forms a Fabry-Perot cavity with the end mirrors (ETM). The light from both arms recombine and interfere with itself destructively at the beam splitter, with the non eliminated light hitting the photodetector (AS port), this in fact is 
the signal we will use to detect gravitational waves. A simple version of the LIGO detector is shown in figure \ref{LIGOsimple}. 

The data taken which measures the strain signal while interferometer is in``lock'' is used to determine if a gravitational wave has been observed. To acquire ``lock'' means to take the interferometer from a state where the mirrors move freely to a state where the light is resonant in the arm cavities and the power-recycling cavity, and the interference at the AS port is kept at the dark point. A more detailed explanation of the LIGO interferometers can be found in \citep{LIGOdetector} and how to ``lock'' the interferometer in \citep{LIGOlock}. The data taken in the locked state and when the feedback systems keep the detector in a ``low noise'' mode is referred to as  ``science data''. This strain signal, \textit{s(t)}, is defined as:
\begin{equation}\label{e:strain} 
s(t) = \frac{\Delta L_x - \Delta L_y}{L}
\end{equation}
where  \textit{L$_{x,y}$} are the arm lengths on the interferometer. This signal is limited by the amount of noise in the interferometer, $n(t)$, which will be described in the following section.

\subsection{Noise Sources}
\label{noisesources}
\begin{figure}[!hbp|t]
\begin{center}
   \centering
   \includegraphics[height=8cm,width=10cm,angle=0]{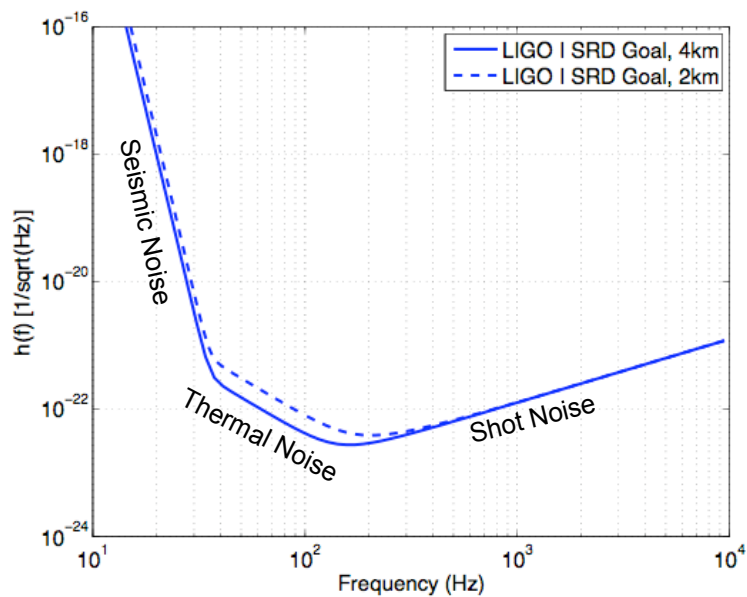} 
   \caption{Noise limits for the LIGO interferometer design \citep{LIGOsensitivity}.}
   \label{LIGOdesignnoise}
\end{center}
\end{figure}

There are several types of noise sources which could affect LIGO and limit LIGO sensitivity. These noise sources contribute to the measured output, denoted as $h(f)$. We show in figure \ref{LIGOdesignnoise}, the three main noise sources which determine LIGO's design sensitivity. These are \textit{seismic noise}, \textit{thermal noise}, and \textit{laser shot noise}. Seismic and thermal noise affect the interferometer by altering the differential change in the arm lengths measured in the anti-symmetric port, and thus the strain signal defined in equation \ref{e:strain}. 

\textit{Seismic noise} limits the strain sensitivity at frequencies f $\lesssim$ 40 Hz. There are several constituents that contribute to the measured seismic noise, including earthquakes (0.03 $\lesssim$ f $\lesssim$ 0.1 Hz), ocean waves (0.1 $\lesssim$ f $\lesssim$ 0.35 Hz), and noise made by the everyday living of people (1.0 Hz $\lesssim$ f $\lesssim$ 3 Hz). This noise is reduced by combination of passive and active isolation techniques  \citep{seismicnoiseredcution}.

\textit{Thermal noise} may be the limit in the 40 Hz $\lesssim$ f $\lesssim$ 200 Hz regime and is due to thermal fluctuations of the individual elements that compose the test masses (fused silica), including the coating on the mirrors, the substrates, and the steel wires that hold the optics in place. 

\textit{Laser shot noise} affects the strain signal sensed in equation \ref{e:strain} and dominates for  f  $\gtrsim$ 200 Hz. This is due to fluctuations in the stored power in the interferometer arm Fabry-Perot cavities from fluctuations in the vacuum field, alternatively interpreted as photon counting statistics (shot noise) on the light. 

The noise sources described above limit the overall sensitivity of LIGO. Actual LIGO detectors are limited by these and a variety of other noise sources, but the overall measured noise currently similar to the LIGO design goal. This is illustrated in figure \ref{LIGOdesignnoise}. 

A more detailed description of all noise sources in the LIGO interferometers can be found in \citep{RanaThesis}. 

\subsection{Science Runs}
\label{scienceruns}

As described in section \ref{design}, science data is taken when the interferometer is in a locked state and is in a low noise mode. Having both LIGO sites operating in a locked state results in times where coincident data can be taken. Once the LIGO sites achieved a sensitivity that was better than previous bar detectors, coincident data taking began, which alternated with commissioning to improve the sensitivity . This was organized into long periods of time of coincident data taking operation, known as science data taking runs, or science runs. The LSC has analyzed the data for gravitational waves in the data taken from these science runs offline. Up to the writing of this thesis, there have been 4 completed science runs with a current science run S5: 
\begin{enumerate}
\item \emph{S1} August 23, 2002 to September 09, 2002
\item \emph{S2} February 14 2003 to 14 April 14 2003
\item \emph{S3} October 31 2003 to January 09 2004
\item \emph{S4} February 22, 2005 to March 24, 2005
\item \emph{S5} Began November 4, 2005 and is currently in progress.
\end{enumerate}

\begin{figure}[!hbp|t]
\begin{center}
   \centering
   \includegraphics[height=8cm,width=10cm,angle=0]{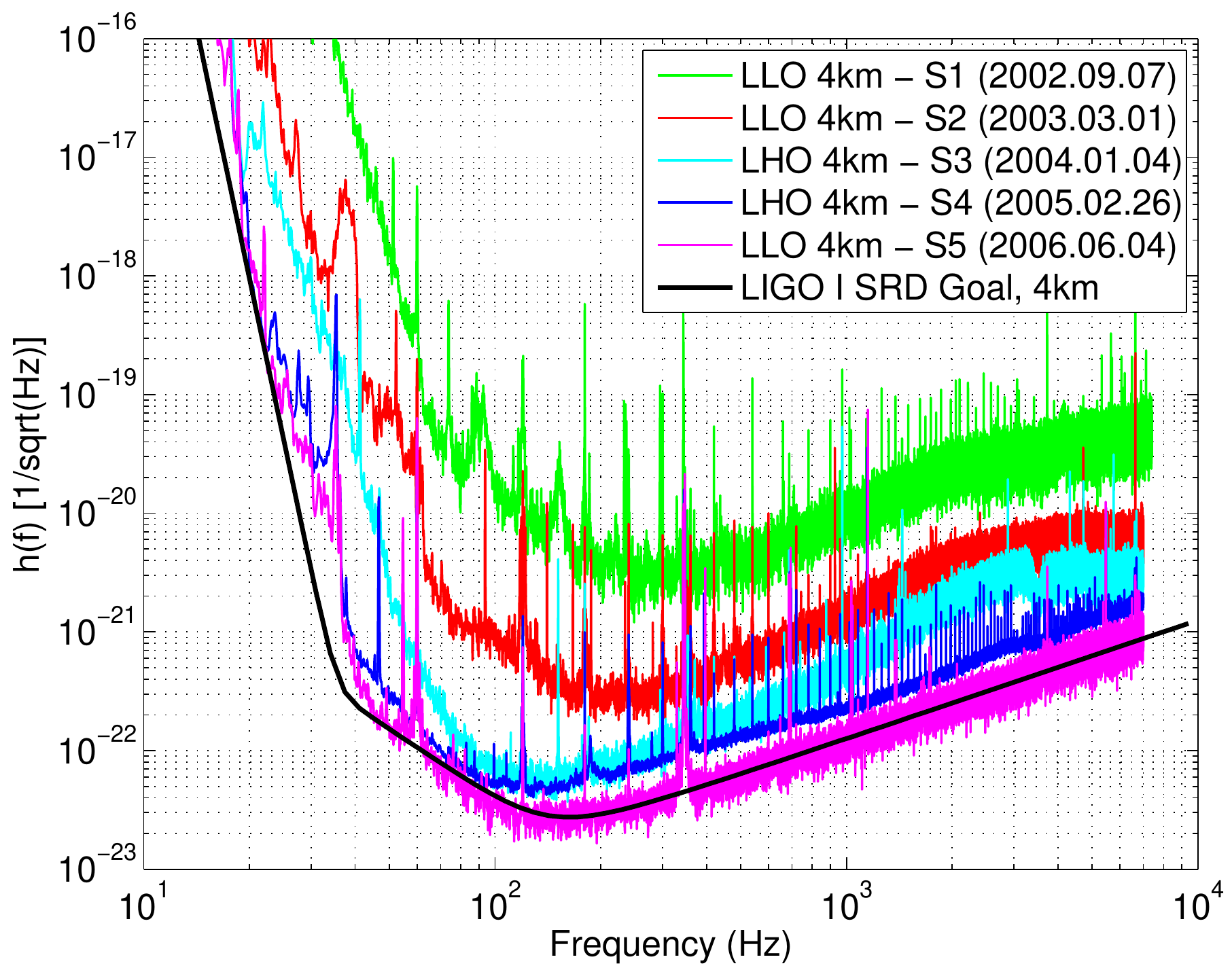} 
   \caption{Best strain sensitivities for the LIGO 4km Interferometers \citep{LIGOstrainBest}.}
   \label{LIGOstrainBest}
\end{center}
\end{figure}

Through S4, no direct evidence of gravitational waves have been detected \citep{S1_1, S1_2, S1_3, S1_4, S2_1, LIGOS2bbh, S2_3, S2_4, S2_5, S2_6, S2_7, S2_8, S2_9, S2_10, S3_1, S3_2, S4_1, S4_2, LIGOS3S4all, S3S4_2, S4_3, S4_4, S4_5}. In S5, the experiment reached the goal in sensitivity where we are now surveying hundreds of galaxies. The improvement in strain sensitivity over the course of the science runs is shown in figure \ref{LIGOstrainBest}.

\renewcommand{\baselinestretch}{1}
\CHAPTER{\singlespace Data Analysis for Coalescing Binary Systems}
\label{chapter3}
\doublespace
In this chapter I introduce the fundamentals of gravitational wave data analysis for searching  coalescing binary systems. We begin with the theory of matched filtering, followed by a description of the method used in a multi-detector search by the LSC in a multi-detector search, the ``inspiral pipeline''. The chapter also discusses methods to tune inspiral searches, and vetoes used, both instrumental and signal-based. 
\doublespace
\section{Optimal Matched Filtering}
\label{matchedfiltering}
In searching for gravitational waves with well understood waveforms, such as ``chirp'' waveforms from inspiralling neutron stars, matched filtering \citep{Cutler1994,Balasubramanian1996} is the optimal detection strategy in Gaussian noise. The measured detector's stain amplitude is given by
\begin{equation}
\label{e:detector_output}
s(t) = n(t) + h(t),
\end{equation}
where $n(t)$ is the detector noise (assumed in this chapter to be Gaussian distributed) and $h(t)$ is the gravitational wave signal. An interferometric detector, such as LIGO (section \ref{LIGO}), is sensitive to a linear combination of the two gravitational wave polarizations (equations \ref{hplus}, \ref{hcross}), where the gravitational wave signal has the form:
\begin{equation} 
\label{h_t}
h(t) = F_{+}h_{+}(t) + F_{\times}h_{\times}(t)
\end{equation} 
with 
\begin{equation} 
\label{F_plus}
 F_{+} = -\frac{1}{2}(1+\cos^{2}\theta)\cos2\phi
 \end{equation}
 \begin{equation} 
\label{F_cross}
F_{\times} = \cos\theta\sin2\phi
\end{equation}  
where F$_{+}$ and F$_{\times}$ are the two antenna response functions of the detector, $\theta$, and $\phi$ are the spherical coordinates of the sources sky position with respect to axes defined by LIGO's arms. 

\begin{figure}[!hbp|t]
\begin{center}
   \centering
   \includegraphics[height=7cm,width=8cm,angle=0]{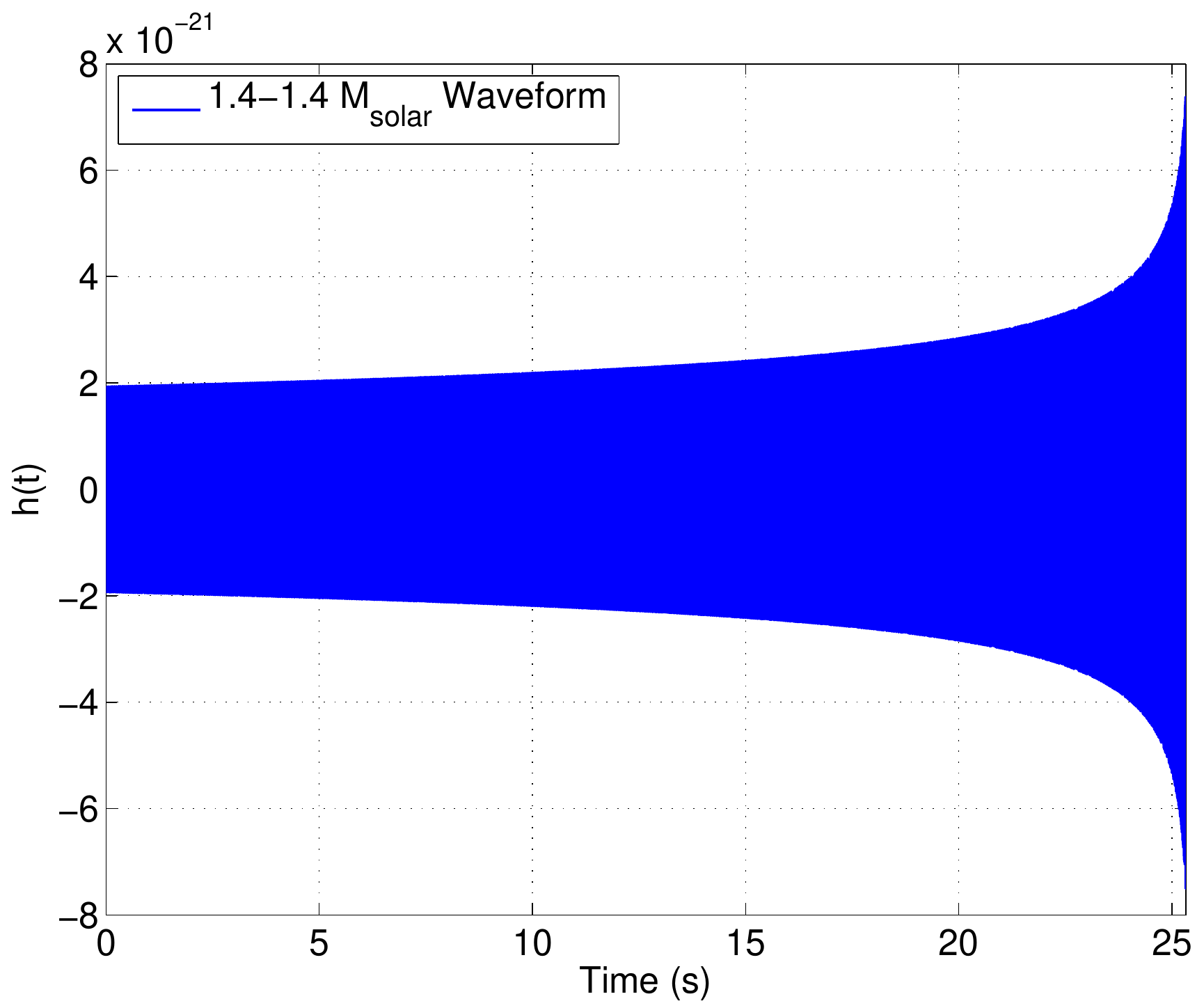} \\
      \includegraphics[height=7cm,width=8cm,angle=0]{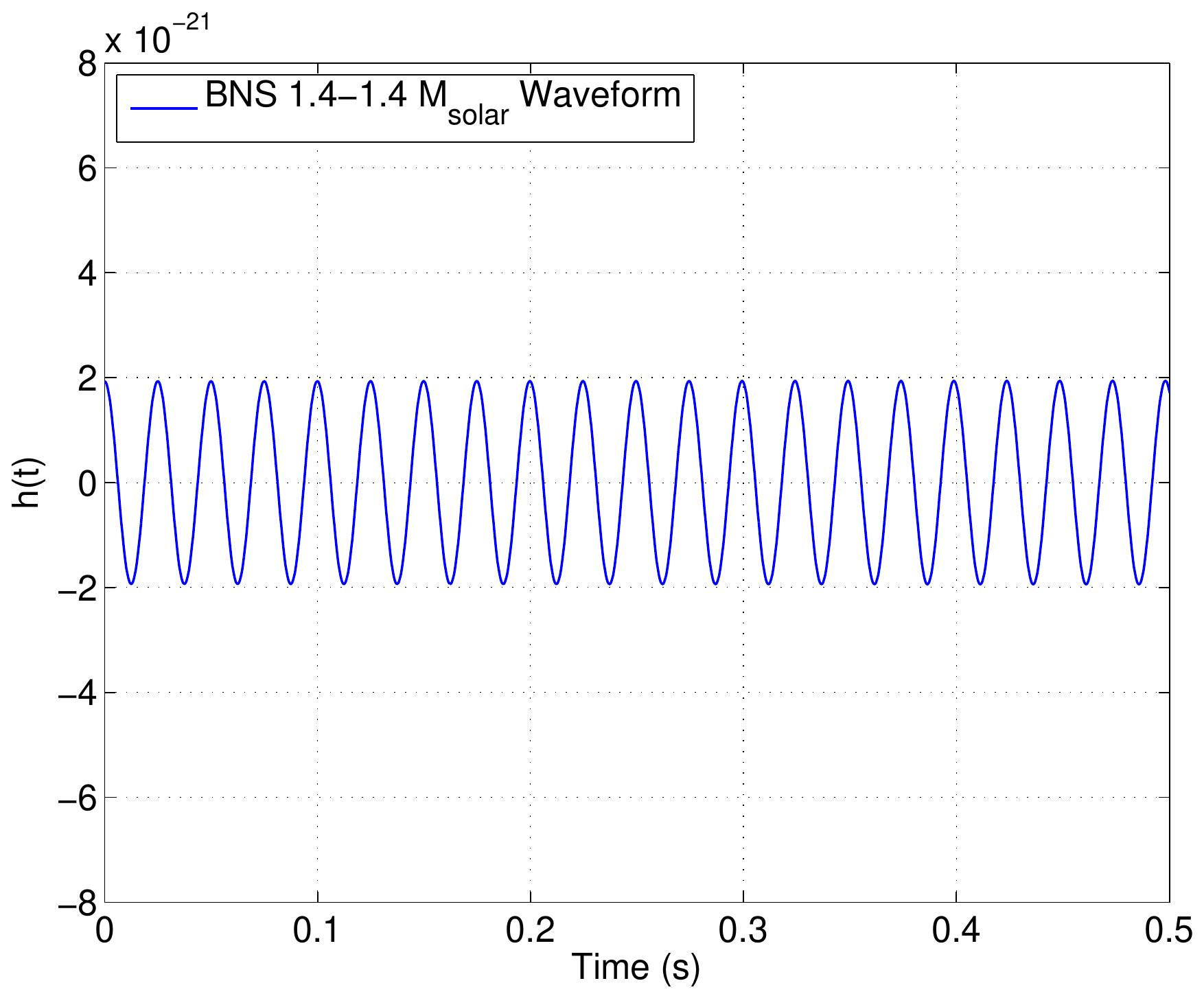} 
   \caption{\singlespace An example chirp waveform for a 1.4$-$1.4 $M_{\odot}$ coalescing binary system, $f$ = 40Hz at $t$ = 0s, $f$ = 1532Hz at $t$ = 25s. A zoom of the first 0.5 seconds is shown in the bottom panel.}
   \label{BNSChirp}
\end{center}
\end{figure}

We transform the template, $h(t)$ (equations \ref{hplus}, \ref{hcross}, \ref{h_t}, \ref{F_plus}, \ref{F_cross}), into the frequency domain using the the second order post-Newtonian \emph{stationary phase approximation}\citep{Droz1999}
\begin{equation} 
\label{e:h_c}
  \tilde{h}_c(f) =  \frac{2GM_\odot}{c^{2}r}\left( \frac{5\mu}{96M_\odot}\right)^{1/2}\left(\frac{M}{\pi^2M_\odot}\right)^{1/3}f^{-7/6}\left(\frac{GM_\odot}{c^3}\right)^{-1/6}e^{i\Psi(f;M,\eta)},
\end{equation} 
\begin{equation} 
\label{e:h_s}
  \tilde{h}_s(f) =  i\tilde{h}_c(f)
  \end{equation} 
where $f$ is the frequency of the gravitational wave, $M$ is the total mass of the binary system, $M$ =  ($m_{1} + m_{2}$), $\mu$ is the reduced mass of the system, $\mu$ = $m_1m_2 / (m_1+m_2)$, and $\eta$ is the ratio $\mu / M$. The template $\tilde{h}(f)$ (units of 1/Hz) is a linear superposition of $\tilde{h}_{c}(f)$ and $\tilde{h}_{s}(f)$, defined in equations \ref{e:h_c}, \ref{e:h_s}. The phase evolution of the chirp signal, $\Psi(f;M,\eta)$, is given by
\begin{eqnarray} 
\label{e:Psi}
  \Psi(f;M,\eta) & =  & 2\pi ft_{c} - 2\phi_{0} - \pi/4 + \frac{3}{128\eta}\Big[x^{-5} + \left(\frac{3715}{756} + \frac{55}{9}\eta \right)x^{-3} - 16\pi x^{-2}  \nonumber\\ && + \left(\frac{15293365}{508032} + \frac{27145}{504}\eta + \frac{3085}{72}\eta^2 \right)x^{-1}\Big ], \\
x & = & { \left( \frac{\pi M f G}{c^3} \right)}^{-1/3}
\end{eqnarray} 
where $\phi_{0}$ is the coalescence  phase, and $t_{c}$ is the coalescence time of the binary. A combination of equations \ref{e:h_c}, \ref{e:h_s}, \ref{e:Psi}, and 3.5 yields the form of the chirp signal used in the search. An example chirp signal for a 1.4$-$1.4 $M_{\odot}$ binary with a starting frequency of 40Hz is given in figure \ref{BNSChirp}.

    Optimal performance is achieved by filtering the data $s(t)$ in the frequency domain weighted inversely by the power spectral density of the noise. The complex output of the Wiener optimal filter is given by:
\begin{equation}
\label{e:optimal_filter}
  z(t) =  4\int_{0}^{\infty} \frac{\tilde{h}^\ast(f)\tilde{s}(f)}{S_{\rm n}(f)}\,e^{2\pi ift} df
\end{equation}
where $\tilde{s}(f)$ is the Fourier transformed data of the detector and $S_{\rm n}(f)$ is the power spectral density of the detector noise. The goal is to locate the maxima of the output of the matched filter, $\|$$z(t)$$\|$, over the arrival time and phase, comparing the maximum value with the expected distribution of values in the absence of the signal. Furthermore, the output of the optimal filter for each template has a characteristic amplitude:
\begin{equation}\label{e:sigmasq} 
  \sigma^2 = 4\int_0^\infty \frac{|\tilde{h}(f)|^2}{S(f)} df,
\end{equation} 
we can then normalize the filter output, defining a signal-to-noise ratio ($\rho$):
\begin{equation}\label{e:rho} 
\rho(t) = \frac{|z(t)|}{\sigma}
\end{equation}
A template has a time series $\rho(t)$, while a candidate event (or trigger) is related to a local maximum of $\rho (t)$, and is assigned an SNR:
\begin{equation}\label{e:rho_candidate} 
\rho = \frac{|z_{\rm max}|}{\sigma}
\end{equation}
Given a candidate with a signal-to-noise ratio (SNR) $\rho$, we can also infer another physical characteristic of the system, the effective distance:
\begin{equation}\label{e:effdist}
  D_{\mathrm{eff}} = (\rm 1Mpc)\frac{\sigma}{\rho},
\end{equation} 
which is the distance at which an optimally oriented binary inspiral would produce the observed signal-to-noise ratio. D$_{\rm eff}$ is related to the physical distance of the binary, r, by the orientation of the binary with respect to the plane of the detector and the detectors two antenna response functions by \citep{Thorne87}:
\begin{equation} 
D_{\rm eff} = \frac{r}{\sqrt{F_{+}^{2}(1+\cos^{2}\iota)^{2}/4+F_{\times}^{2}\cos^{2}\iota}}
\end{equation} 
A true inspiral signal would have a narrow peak in $\rho$$(t)$ at the coalescence time, $t_c$, the end of the template used, or the time at which the two stars begin to merge. If no gravitational wave is present, all we would have is stationary detector noise and $\langle \rho^2(t) \rangle$ = 2,  since $\rho(t)$ is a random quantity, $\chi^2$ distributed with 2 degrees of freedom. 

Instrumental glitches in the data can lead to large $\rho$, even in the absence of a signal: these are called ``false alarms''. Further tests are needed to differentiate real signals from false alarms; these include coincidence between detectors, the $\chi^2$ veto.  and the r$^2$ test, which will be further investigated in \ref{chapter4}.

A more detailed explanation of how matched filtering is implemented in LIGO CBC searches can be found in \citep{Duncan_thesis}.

\renewcommand{\baselinestretch}{1}
\section{Searching for Gravitational Waves from Coalescing Binary Systems: The Inspiral Pipeline}
\label{pipeline}
\doublespace
The search for gravitational waves from binary systems in a multi-detector search done by the LSC uses several steps, collectively called the ``inspiral pipeline''. The pipeline begins with the raw data from a set of interferometers and ends with a number of candidate events, which have passed a multitude of tests. These include signal thresholds, multi-tier coincidence, signal based vetoes, and data quality cuts. The pipeline itself can be run with simulated signals, either physically injected into the interferometer (hardware injections) or into the data itself (software injections), and provide a great test of the efficiency to detect inspiral waveforms. 

   The inspiral search pipeline can be divided into the following stages. 
\begin{enumerate}
\item \emph{Data Collection}:  For the LIGO searches, we use a channel named DARM$\_$ERR. This channel is sampled at 16384 Hz and is the error signal of the feedback loop, controlling $L_x$ - $L_y$, keeping the differential arm length (equation \ref{e:strain}) constant.
\item \emph{Template Banks}: The template bank generation begins by reading in the DARM$\_$ERR data, down-sampling it to 4096 Hz, and applying a high-pass filter. The data is divided into 2048 second ``blocks''. These ``blocks'' are further divided into 256 second segments, which are then overlapped with each other by 128 seconds, illustrated in figure \ref{chunk}. This is done to insure no corrupted output is included into the analysis due to edge effects of the Fast-Fourier-Transform. The average power spectral density of the noise, $S(f)$, is also calculated, as well as the instrumental response function, which is used to calibrate the data and the power spectrum. Once this is complete, the template waveforms are generated for each ``block" of data. Note that we require more than one template to scan to full parameter space, therefore using a bank of templates. The number of templates in a bank is governed by allowed mismatch in the signal with the template waveform, which ensures we lose $<$ 3$\%$ in SNR of the signal. In the S4 searches \citep{LIGOS3S4all}, the template banks per block of data was 3500 for BNS, 4500 for PBH, and 1200 for BBH.
\item \emph{Matched Filtering I}: As described previously in section \ref{matchedfiltering}, the templates generated from step 2 are run through each block of data. The SNR time series, ($\rho(t))$, is calculated for each template in each segment, where a SNR threshold, $\rho^*$, is applied. Inspiral ``candidates'' are kept if $\rho$ $>$ $\rho^*$.The SNR threshold ($\rho^*$) is chosen low enough to allow a background of false alarms. 
\item \emph{First Coincidence}: At this stage, the inspiral triggers from multiple interferometers are combined, and we look for triggers coincident in time and mass within some accuracy. The triggers that pass this test are kept, while the others are discarded. Demanding coincidence between interferometers significantly lowers false alarms. 
\item \emph{Triggered Template Bank}: A new template bank is generated using only the parameters of triggers which survived the previous step. Their mass parameters are used to create the new bank. 
\item \emph{Matched Filtering II}: The templates generated from stage 5 are run through the data again.  
\item \emph{Vetoes}: Several instrumental (sec.\ref{Vetoes}) and signal-based (sec.\ref{chisq}, sec.\ref{rsqtest}) vetoes are applied. 
\item \emph{Second Coincidence}: We look again for coincidences between candidates left after in step 7. These are our candidate events. 
\end{enumerate}

  The pipeline is run on a given data set using Directed Acyclic Graph (DAG), which describes the workflow, and is executed using the Condor High Throughput Computing System \citep{condor}, which manages the execution of tasks. The DAG reads in a configuration file which contains locations of executables, science data lists, and search parameters. The output of each stage of the pipeline after the data collection are in the extensible markup language (XML) format, which stores in a tabular format the parameters and information about the triggers. The software used to perform all steps of the analysis and to construct the DAG is in the package \emph{LALAPPS}\citep{LAL}.
   
\begin{figure}[!hbp|t]
\begin{center}
   \centering
   \includegraphics[height=15cm,width=11cm,angle=0]{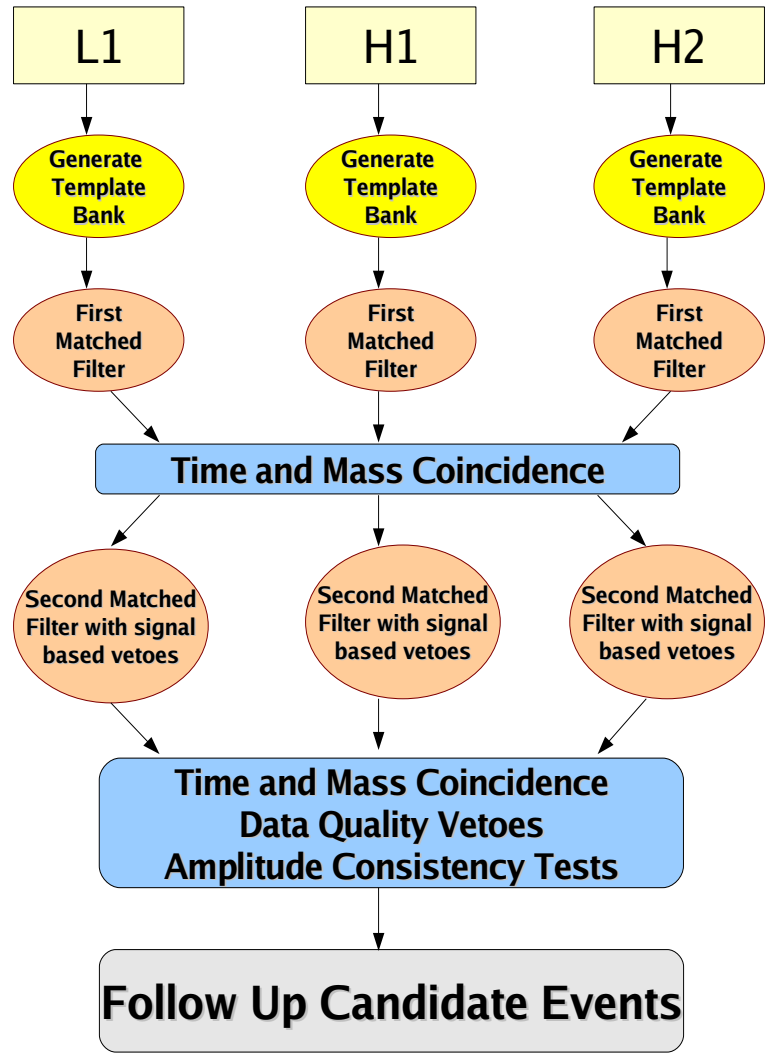} 
   \caption{Illustration of the Inspiral Coincidence Analysis Pipeline \citep{hipepipe}.}
   \label{hipepipeline}
\end{center}
\end{figure}

\begin{figure}[!hbp|t]
\begin{center}
   \centering
   \includegraphics[height=6cm,width=10cm,angle=0]{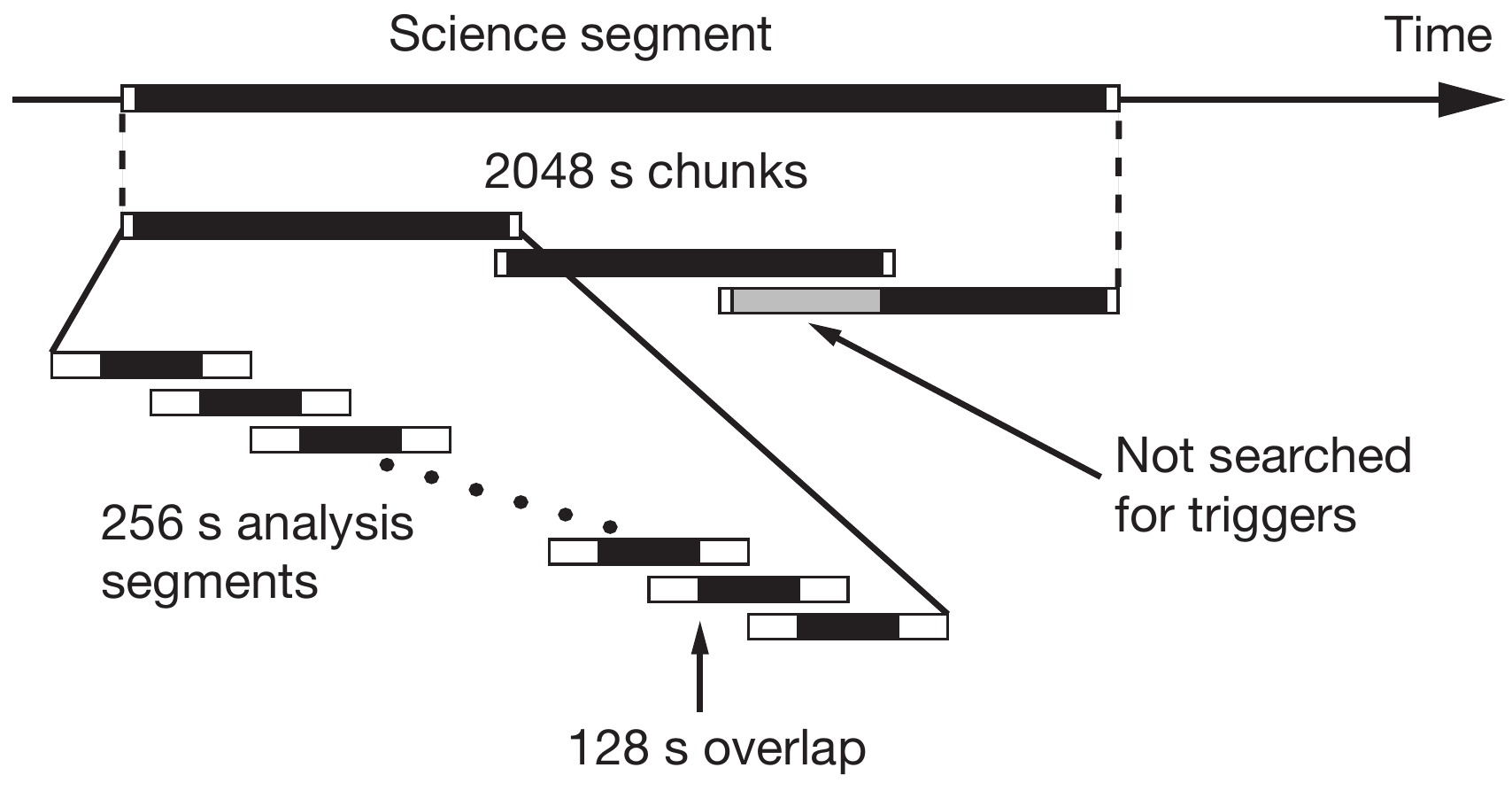} 
   \caption{Illustration of how single interferometer data is divided \citep{S1_2}.}
   \label{chunk}
\end{center}
\end{figure}

  We also repeat the pipeline with time shifts to find the background rate of the false alarms. Time shifts are done by taking one set of interferometer triggers at one detector and shifting them in time, usually on the order of seconds. These time shifted triggers are then run through the first coincidence stage where any coincidence between the time shifted triggers and real interferometer triggers are now false alarms. The pipeline is also repeated for software injections (signals simulated in software) to determine the efficiency of the search. A visual representation of the pipeline described can be found in figure \ref{hipepipeline}.
\section{Tuning the Search for Inspiralling Binaries}
\label{tuning}
Searches for gravitational waves from coalescing binary systems must be optimized using variety of methods in order to increase the likelihood of detecting gravitational waves. In other words, parameters used in different stages of the search pipeline must be tuned. Data from the \emph{playground}, which constitutes roughly 10\% of all the data from a science run taken in 600 second intervals every 6370 seconds, is used to do this. The tuning presented here will be primarily focused on what was used for LIGO's 3$^{rd}$ and 4$^{th}$ science runs for the binary neutron star (BNS) and primordial black hole (PBH) searches. 

A first step in tuning our searches is allowing a time coincidence window taking into account the precision in recovering the gravitational waveform's inferred coalescence time, $t_c$, and the light travel time between detector sites. The precision is measured by injecting simulated gravitational waves into the archived data and running matched filtering. Since we know what the parameters of the injected waveforms are, we can take the output of the matched filters to see how well the detected parameters agree with the injected parameters. We histogram the difference between injected and recovered $t_c$. The width of the histogram gives the timing accuracy. The S3 BNS search used a window of 4 milliseconds to find an injection in a single detector, for coincidence between sites we also include the light travel time, which is on the order of 10 milliseconds.
   
Other coincidence windows are tuned. For non-spinning waveforms, the masses are the main parameters, and we can therefore choose a coincidence test based upon the chirp mass. The chirp mass is given by:
\begin{equation}
\label{e:chirpmass}
\mathcal{M}_c = \frac{(m_1m_2)^{3/5}}{(m_1+m_2)^{1/5}},
\end{equation}
and the symmetric mass ratio of the binary is:
\begin{equation}
\label{e:eta}
\eta = \frac{m_1m_2}{(m_1+m_2)^2}=\frac{\mu}{\mathcal{M}_c}.
\end{equation}
Precision in the recovered chirp masses depends on the variance in the detector noise (equation \ref{e:sigmasq}) and discreteness of the template bank. For the S3 and S4 PBH and BNS searches, we chose a window for both $\eta$ and $\mathcal{M}_c$ based upon histograms the injected and detected parameters. 

A third test can be only done in the case of the two co-located IFO's, such as H1 and H2 by tuning the precision in the amplitude, or the effective distance, defined in equation \ref{e:effdist},  for candidates found in both IFO's. The precision in effective distance is given as
\begin{equation}
\frac{2 |{\rm H1 } D_{\rm eff} - {\rm H2 } D_{\rm eff}|}{{\rm H1 } D_{\rm eff}+ {
\rm H2 } D_{\rm eff}} 
< \kappa  
\end{equation}
where $\kappa$ is the value tuned with the software injections. In S3, the sensitivity of of H1 and H2 differed significantly, and the test used was:
\begin{equation}
\label{e:kappa}
\frac{|{\rm H1 } D_{\rm eff} - {\rm H2 } D_{\rm eff}|}{{\rm H1 } D_{\rm eff}} < \kappa  
\end{equation}
The amplitude consistency check is not done between the Hanford and Livingston detectors since the detectors are not perfectly aligned due to the curvature of the Earth; therefore we could not expect a similar amplitude between sites. 

\section{Vetoes}
\label{Vetoes}
So far I have described the detection strategy used for detecting inspiral signals, the analysis pipeline, and how to tune such a search. I now bring focus to the characterization of vetoes. Conceptually, a veto is a time where by instrumental inspection or some statistical calculation, the data is atypical. There are three classes of vetoes; [1] instrumental vetoes - caused by problems found in the interferometer, [2] statistical vetoes - defined by correlations between the gravitational wave channel and other channels, and [3] signal based vetoes,  which determine if candidate events are consistent with the waveform. 

\subsection{Instrumental Vetoes}
\label{instrumental_vetoes}

An instrumental veto occurs at a time when the instrument was known to malfunction or is sensing an environmental excitation and the malfunction is known to generate false alarms. In the ideal case, once the times due to all instrumental problems are vetoed, the distribution of remaining false alarms is as predicted for Gaussian data.

The goal in veto investigations is several-fold: it should provide a high efficiency of dismissing false signals (high veto efficiency), while keeping the amount of science mode time used at a minimum (low deadtime), and it should have a large percentage of veto triggers that eliminate one inspiral event or more. It is a balance between these factors that we try to achieve, with the more analyzable time we have with Gaussian noise data, the greater the chance of detection.

There are times during any of science runs when environmental disturbances and non-optimal interferometer performance times should be excluded from the analysis, even when the interferometer is in science mode\citep{Nelson}. Large transients in the data causes most, if not all of the templates in the bank to ring off, and thus create false alarms, even in coincidence. Members of the LSC doing detector characterization develop data quality (DQ) flags. Some of the tuning is done with candidates from the first matched filtering stage of the analysis pipeline (sec.\ref{pipeline}). These triggers are used as a primer to go further into the IFO performance and look for causes of these occurrences. The result of such work are data quality (DQ) flags, used to let the scientists know what was discovered in the IFO while it was in science mode. A DQ flag can be created for a number of reasons, ranging from earthquakes affecting the instrument, dips of the light intensity in the arms, airplanes passing by, and an unrelated channel triggering. It should be noted that not all DQ flags are created from inspiral triggers, most are in fact created from known problems with the interferometer. Some DQ flags do not produce false alarms in compact binary coalescence (CBC) searches, thus, tuning is required. 

\subsection{Statistical Vetoes}
\label{statistical_vetoes}

Problems in the instrument can be identified in some statistical ways . An example includes looking for correlations between ``kleineWelle triggers'' \citep{Chatterji} on the gravitational wave (GW) channel and other auxiliary channels.  kleineWelle is a wavelet analysis algorithm. An example of such a veto is the ``AS\_I veto'' \citep{Channa}, where times when there are kleineWelle triggers in both AS\_I and AS\_Q, with a ratio that makes the transient a very unlikely gravitational wave. 

\subsection{Signal Based Vetoes}

Finally, there is another category of vetoes which includes signal-based vetoes, which are used to discriminate a real signal from background noise in a given search. An example of this is the $\chi^2$ veto and the r$^2$ veto, which is based upon the time series of $\chi^2$. The following chapter will describe these both in detail.

\renewcommand{\baselinestretch}{1}
\CHAPTER{Methods to Reduce False Alarms in Coalescing Binary System Searches}
\label{chapter4}
\doublespace
This chapter introduces a veto currently implemented in searches of gravitational waves from coalescing binary systems, the $\chi^2$ veto, and introduces a new veto, the r$^2$ test. In the last section, I will describe how the test is incorporated into the inspiral pipeline used in LSC searches. The results presented in this chapter are my own while collaborating with the CBC group.
\section{The $\chi^2$ Veto}
\label{chisq}
\begin{figure}[!hbp|t]
\label{chisqvisual}
\begin{center}
   \centering
   \includegraphics[height=6cm,width=8cm,angle=0]{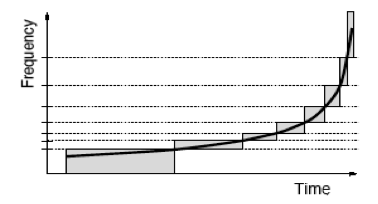} 
   \caption{\singlespace A visual representation of $\chi^2$ statistic. The solid black line represents the chirp waveform, while the gray boxes illustrate each time-frequency bin, p = 8 in this example \citep{PShawhan}.}
\end{center}
\end{figure}

As described in Sec.\ref{matchedfiltering}, matched filtering provides the optimal detection strategy for detecting inspiral waveforms in Gaussian noise. Noise artifacts present in non-Gaussian noise data generate a signal-to-noise ratio ($\rho$) large enough to exceed the given search threshold ($\rho^*$) resulting in a false alarm. In order to further reduce these transient signals from being misidentified as real signals, we check whether the waveform has the expected accumulation of $\rho$ in several time-frequency bins, illustrated in figure \ref{chisqvisual}. This is analogous to breaking one template into several sub templates $\emph{i}$, where each sub template produces $\rho$$/$$p$ of its parent template's SNR $\rho$.

For each candidate, we define the $\chi^2$ time series as \citep{BAllen}:
\begin{equation}
  \chi^2(t) = \frac{p}{\sigma^{2}}\sum_{l=1}^p |z_l(t)-z(t)/p|^2 = p\sum_{i=1}^p |\rho_i(t)-\rho(t)/p|^2
\end{equation}
where p is the number of $\chi^2$ bins (sub templates), $\rho(t)$ is the matched filter (SNR time series) for the signal, and $\rho_i(t)$ is the SNR time series of the sub templates, \emph{i}. If the noise in the detector was Gaussian noise, we expect $\chi^2$$(t)$ to be  $\chi^2$ distributed with 2p-2 degrees of freedom in the absence of true signals. The reduced $\chi^2$ test threshold ($\chi^{*2}$) for Gaussian noise is:
\begin{equation}
\label{chisqgaussthreshold}
\chi^{*2} <  \frac{\chi^2}{<\chi^{2}>}=\frac{\chi^2}{2p-2}
\end{equation}
where the quantity of the right has an expected value of 1. In LIGO inspiral searches, we re-normalize $\chi^{2}$ to have an expected value $\approx$ 2.
\begin{equation}
\label{rsq}
r^2 =  \frac{\chi^2}{p}
\end{equation}

This quantity is referred to as r$^2$ \citep{rsqref}. For inspiral searches with LIGO data, a discrete template bank is used that causes a potential mismatch between the template waveform and signal ($\sim$ 5\% or less). The mismatch causes the $\chi^2$ to scale with $\rho^2$. The r$^2$ must be modified to account for this. Therefore, we add a non-central parameter $\delta \rho^2$ to r$^2$: 
\begin{equation}
\label{xisquared}
\xi^2 = \frac{\chi^2}{p+ \delta \rho^2} 
\end{equation}
where $\delta$ is the mismatch between the template and the signal. This is called the \emph{weighted} $\chi^2$, or $\xi^2$.  In order allow room for these mismatched signals, we apply a threshold on $\xi^2$, denoted  $\xi^{*2}$. We therefore ask when $\rho$ exceeds  $\rho^*$, that the triggers also have both:
\begin{equation}
\label{e:xistarsquared}
\frac{\chi^2}{p+ \delta \rho^2} < \xi^{*2} 
\end{equation}
Since this is  a new constraint on the outputted triggers, the $\chi^2$ test now becomes a veto, hence the $\chi^2$ veto. 

\begin{figure}[!hbp|t]
\begin{center}
\includegraphics[height=7cm,width=8cm,angle=0]{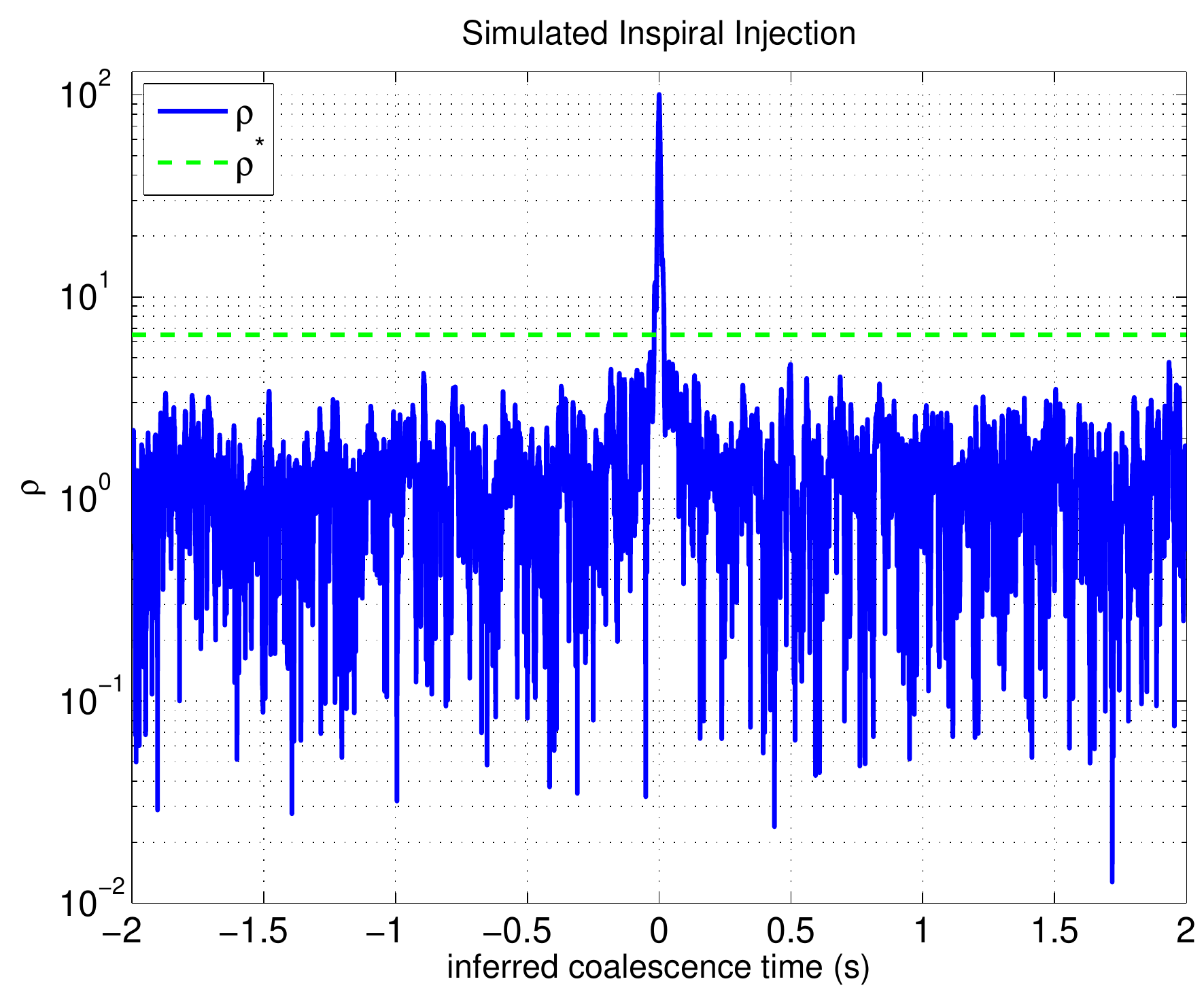} \\
\includegraphics[height=7cm,width=8cm,angle=0]{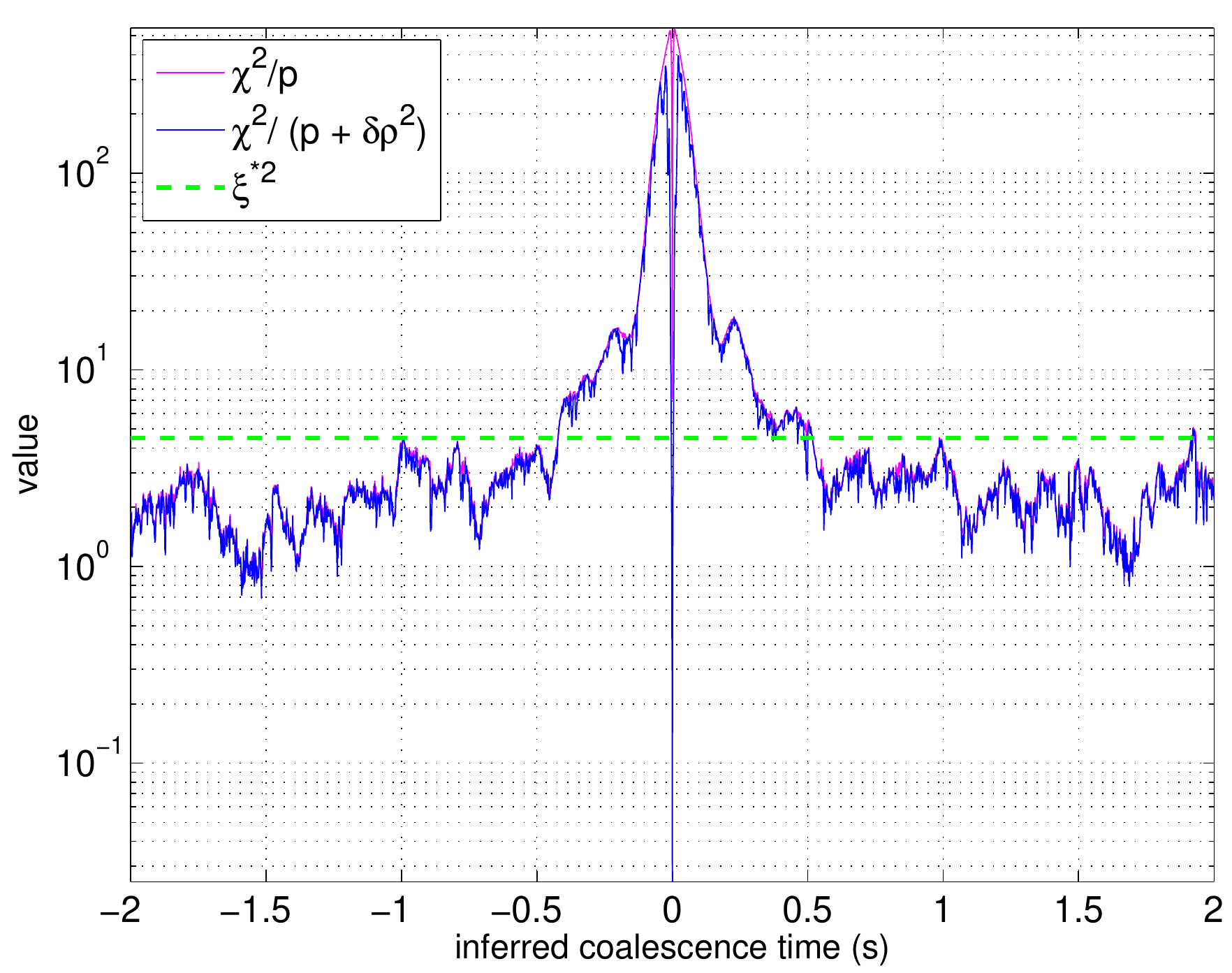} \\
\includegraphics[height=7cm,width=8cm,angle=0]{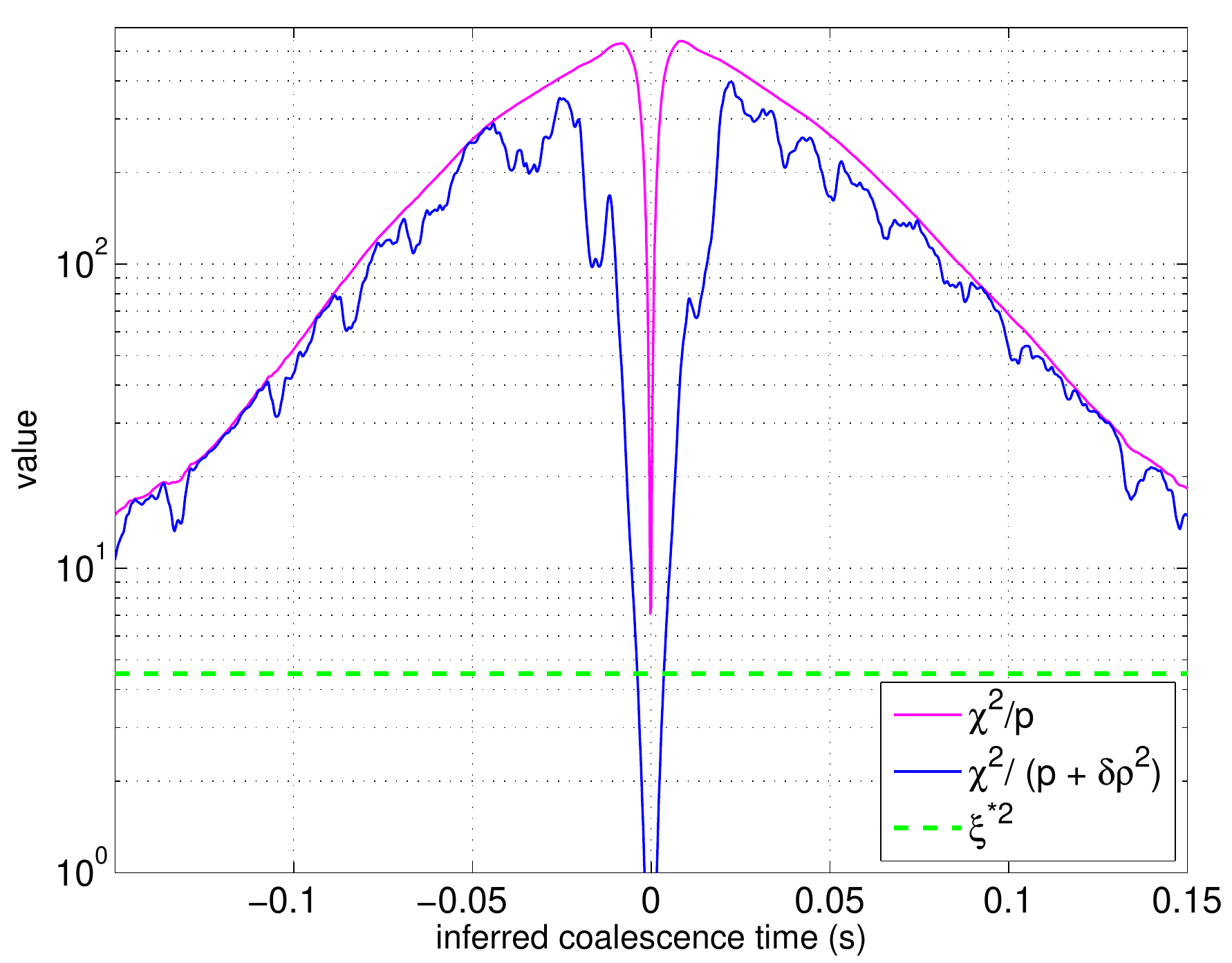} 
\end{center}
\caption{\singlespace A simulated inspiral injection SNR and weighted $\chi^2$ time series using LIGO data. The top panel is a plot of the SNR time series, the middle panel plots the weighted $\chi^2$ time series and $\chi^2$$/$p. The bottom panel is a zoom of the middle panel. The dashed lines represent the set thresholds.}
\label{triginjproof_highSNR}
\end{figure}

An example comparing the output for a high $\rho$ simulated inspiral injection's $r^2$ and weighted $\chi^2$ time series is shown in figure \ref{triginjproof_highSNR}. Notice in the zoomed in bottom panel in figure \ref{triginjproof_highSNR} where the $\xi^{*2}$ threshold for the loud injection would have vetoed the calculated r$^2$. 

\renewcommand{\baselinestretch}{1}
\section{A Test to Further Reduce False Alarms}
\label{rsqtest}

\begin{figure}[!hbp|t]
\begin{center}
   \centering
   \includegraphics[height=8cm,width=10cm,angle=0]{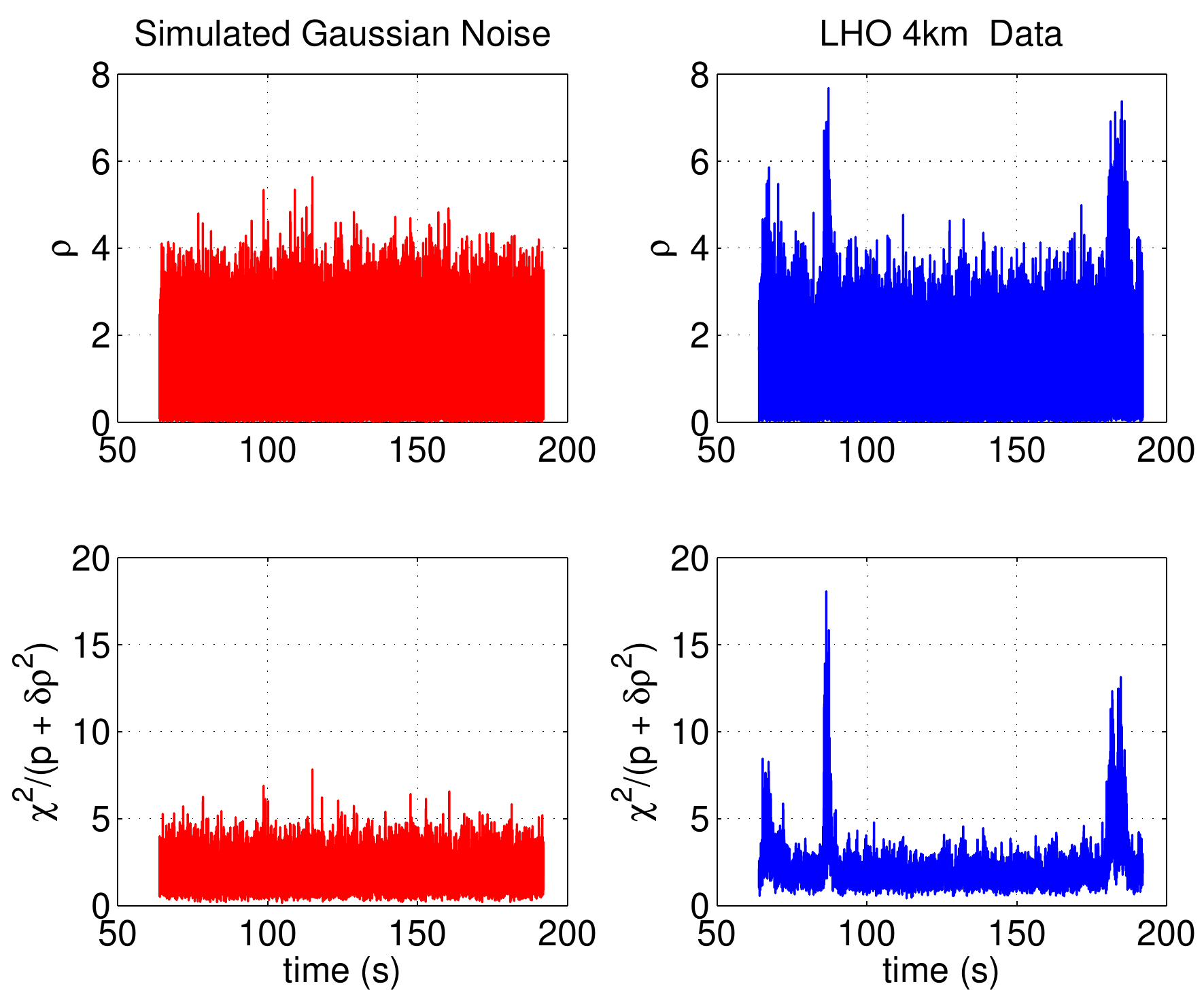} 
   \caption{\singlespace The SNR and weighted $\chi^2$ time series for a 128 second segment of LIGO data and simulated stationary Gaussian noise}
   \label{snrrsqtimeseries}
\end{center}
\end{figure}

In the inspiral pipeline (section \ref{pipeline}), triggers are kept during the second matched filtering stage if they have a SNR above a threshold ($\rho^*$) and the weighted $\chi^2$ below a threshold ($\xi^{*2}$). In Gaussian noise, the probability of these triggers being a false alarm can be estimated. In real data collected by LIGO for example, this estimate is seriously flawed due to the presence of excess noise contributed from instrumental transients. This is exemplified by comparing segment of data's filter output for a single template's $\rho(t)$ and weighted $\chi^2$ time series in simulated Gaussian noise and actual detector noise (LHO 4km data). This is plotted in figure \ref{snrrsqtimeseries}.
\begin{figure}[!hbp|t]
\begin{center}
   \centering
   \includegraphics[height=8cm,width=10cm,angle=0]{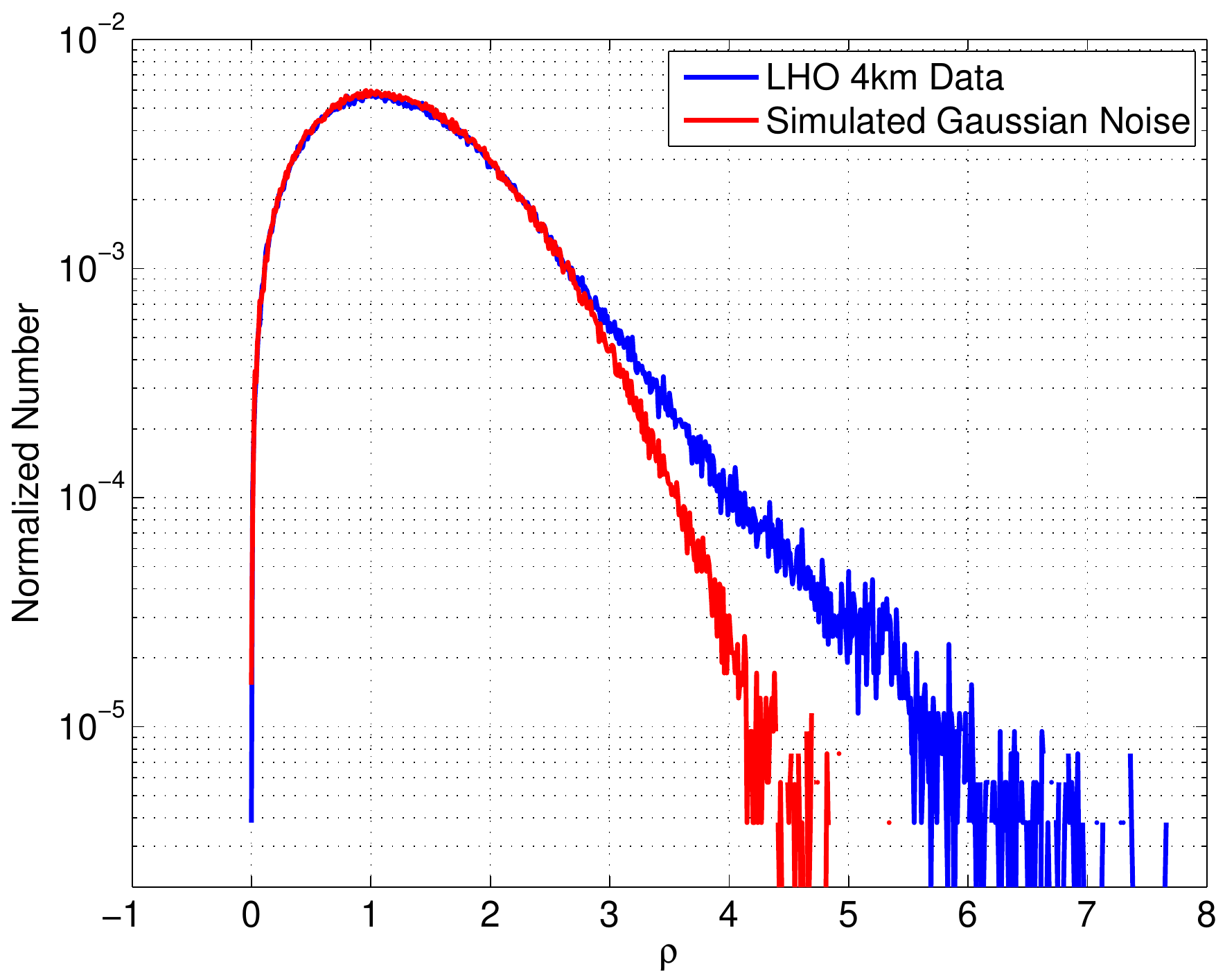} 
    \includegraphics[height=8cm,width=10cm,angle=0]{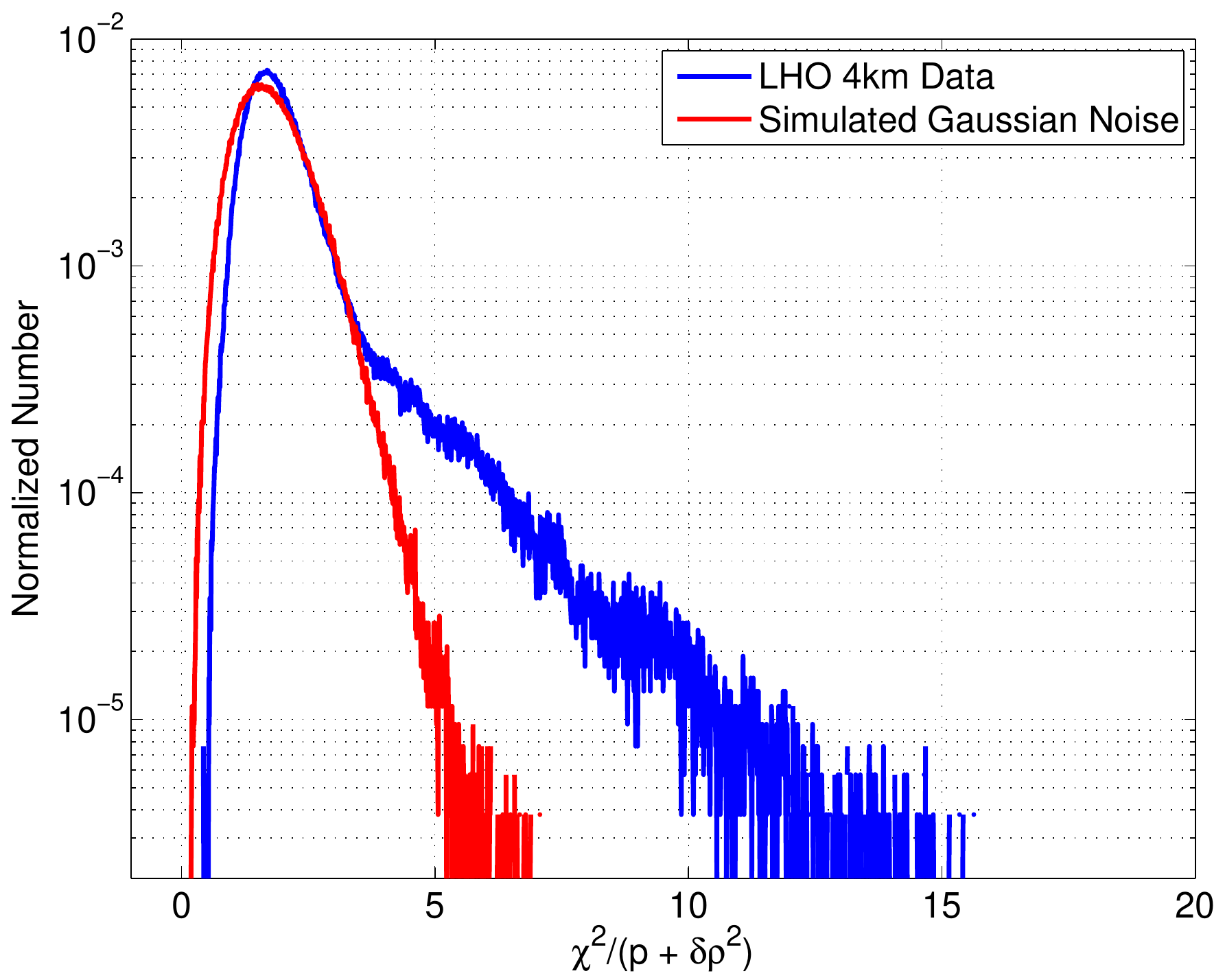} 
   \caption{\singlespace A histogram of figure \ref{snrrsqtimeseries}. The top panel is a histogram of the SNR time series. The bottom panel is a histogram of the weighted $\chi^2$ time series. }
   \label{rsqtimeserieshisto}
\end{center}
\end{figure}
In the simulated Gaussian noise case, both time series stay relatively flat with some small fluctuation, while in the detector noise, we see times when the SNR rises and falls and the weighted $\chi^2$ time series changing by a significantly larger amount. Figure \ref{rsqtimeserieshisto} shows histograms of the SNR and the weighted $\chi^2$ time series. Notice here the SNR time series overlaps in both cases, with the LIGO data having a tail. The weighted $\chi^2$ has a slight shift in overlap, due to the presence of instrumental glitches. We want to develop a method to characterize the excess noise seen in these figures, and therefore create a veto to reduce the background. This can be done based of  the time history of SNR $\rho$$(t)$ or weighted $\chi^2$ time series, or some variant thereof. 

\begin{figure}[!hbp|t]
\begin{center}
   \centering
   \includegraphics[height=8cm,width=10cm,angle=0]{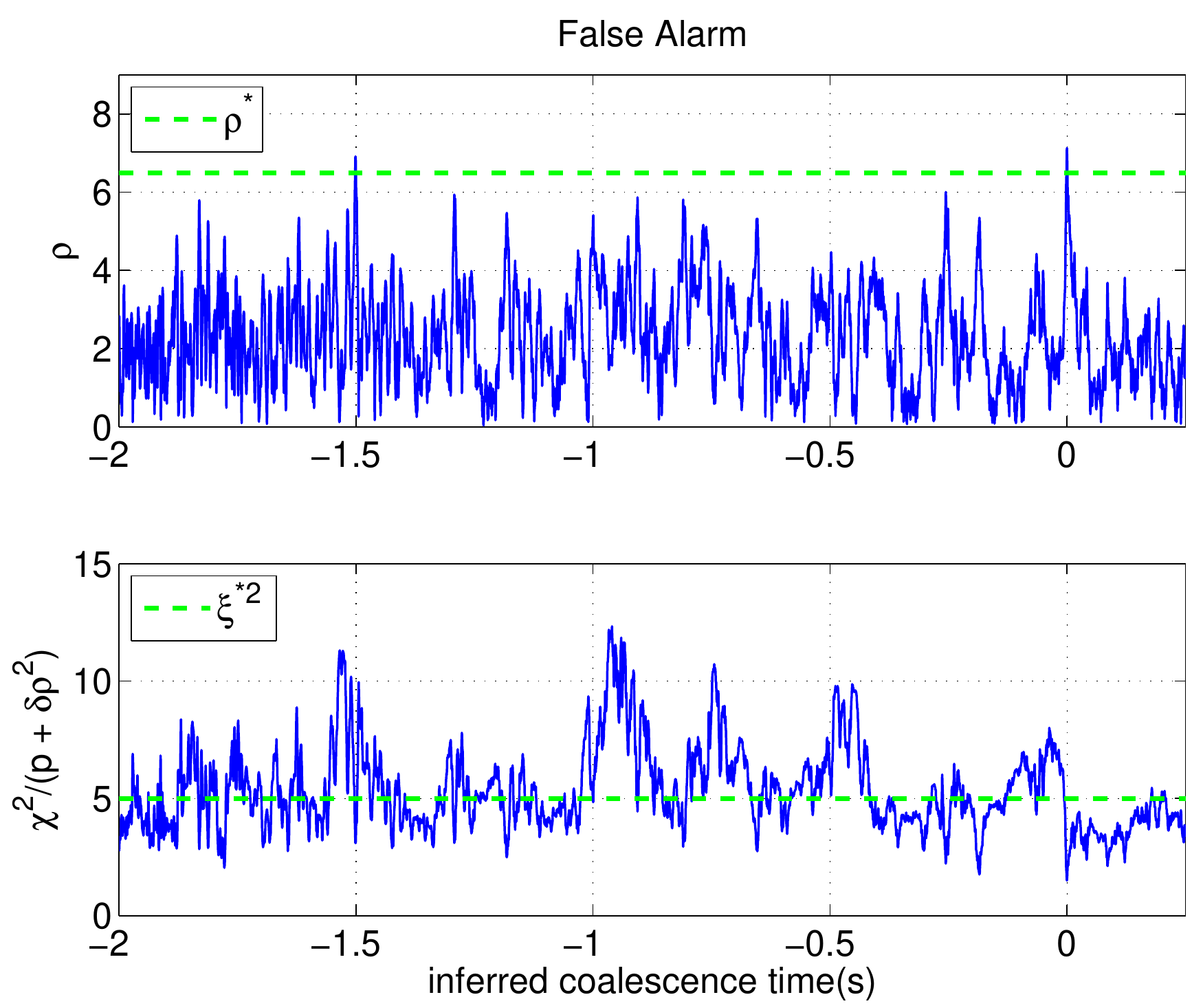} 
   \includegraphics[height=8cm,width=10cm,angle=0]{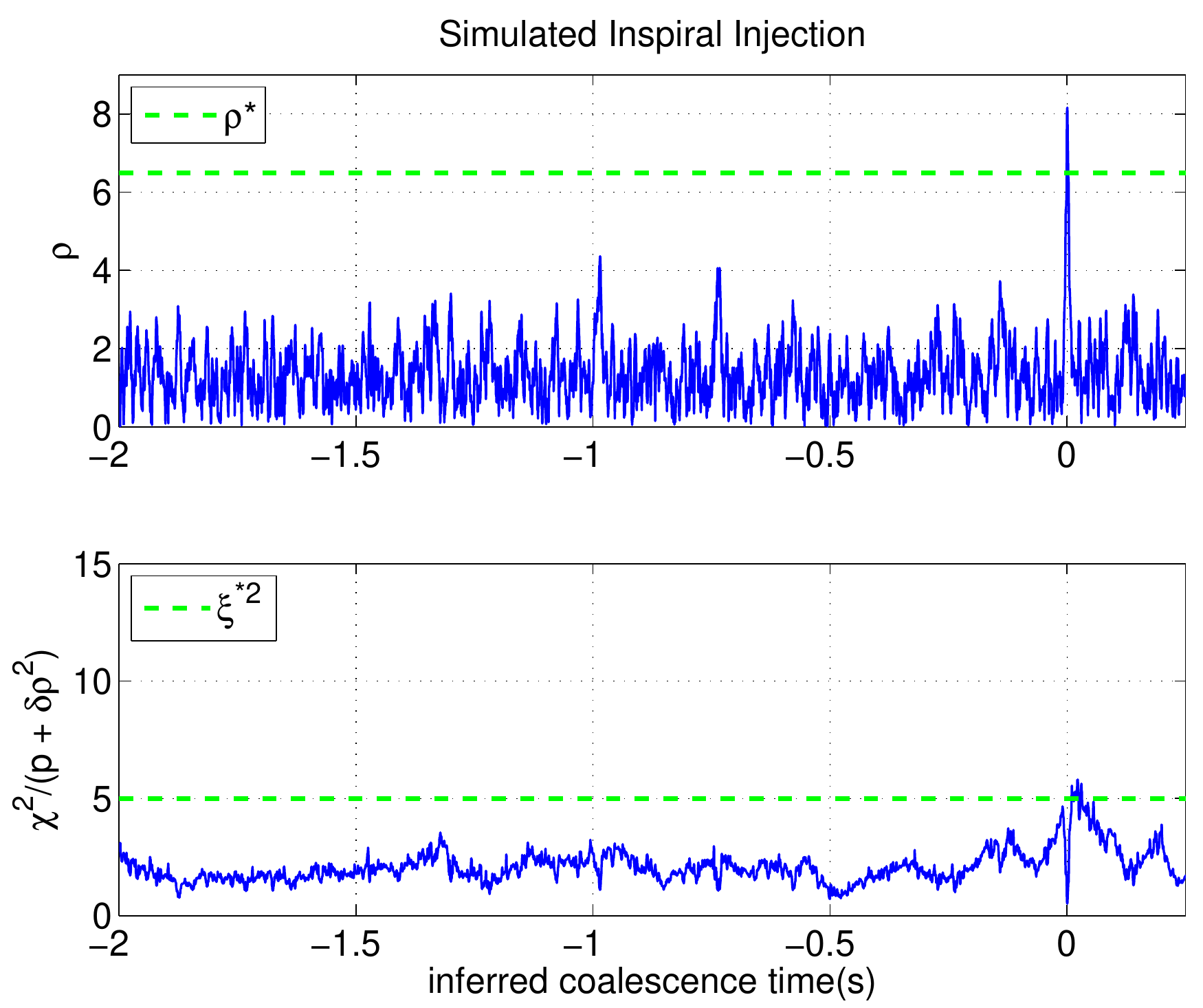} 
   \caption{\singlespace A comparison of SNR and weighted $\chi^2$ time series for a false alarm (top two panels) using LIGO data and a simulated inspiral injection (bottom two panels) in simulated stationary Gaussian noise. The dashed lines represent the set thresholds.}
   \label{triginjproof}
\end{center}
\end{figure}
Take as an example figure \ref{triginjproof}. The time series, $\rho$(t), of the false alarm and simulated inspiral injections behave differently up to the coalescence time. The weighted $\chi^2$ time series has greater fluctuations for the false alarms up to the inferred coalescence time, making it easier to distinguish it from a true signal. Therefore, we choose to use create a test based upon the $\chi^2$ time series as a method to search for excess noise. We use r$^2$ (equation \ref{rsq}) instead of the weighted $\chi^2$ (equation \ref{xisquared}), due to the smooth behavior of r$^2$ and also since the weighted $\chi^2$ has a minor effect until we reach the coalescence time of the simulated inspiral injection  as shown in the bottom panel of figure \ref{triginjproof_highSNR}. The test will measure the consistency of the candidate signal using r$^2$(t) by scanning a given amount of time prior to the coalescence time at a given threshold. This test is otherwise known as the r$^2$ test.  
   
\begin{figure}[!hbp|t]
\begin{center}
   \centering
   \includegraphics[height=8cm,width=10cm,angle=0]{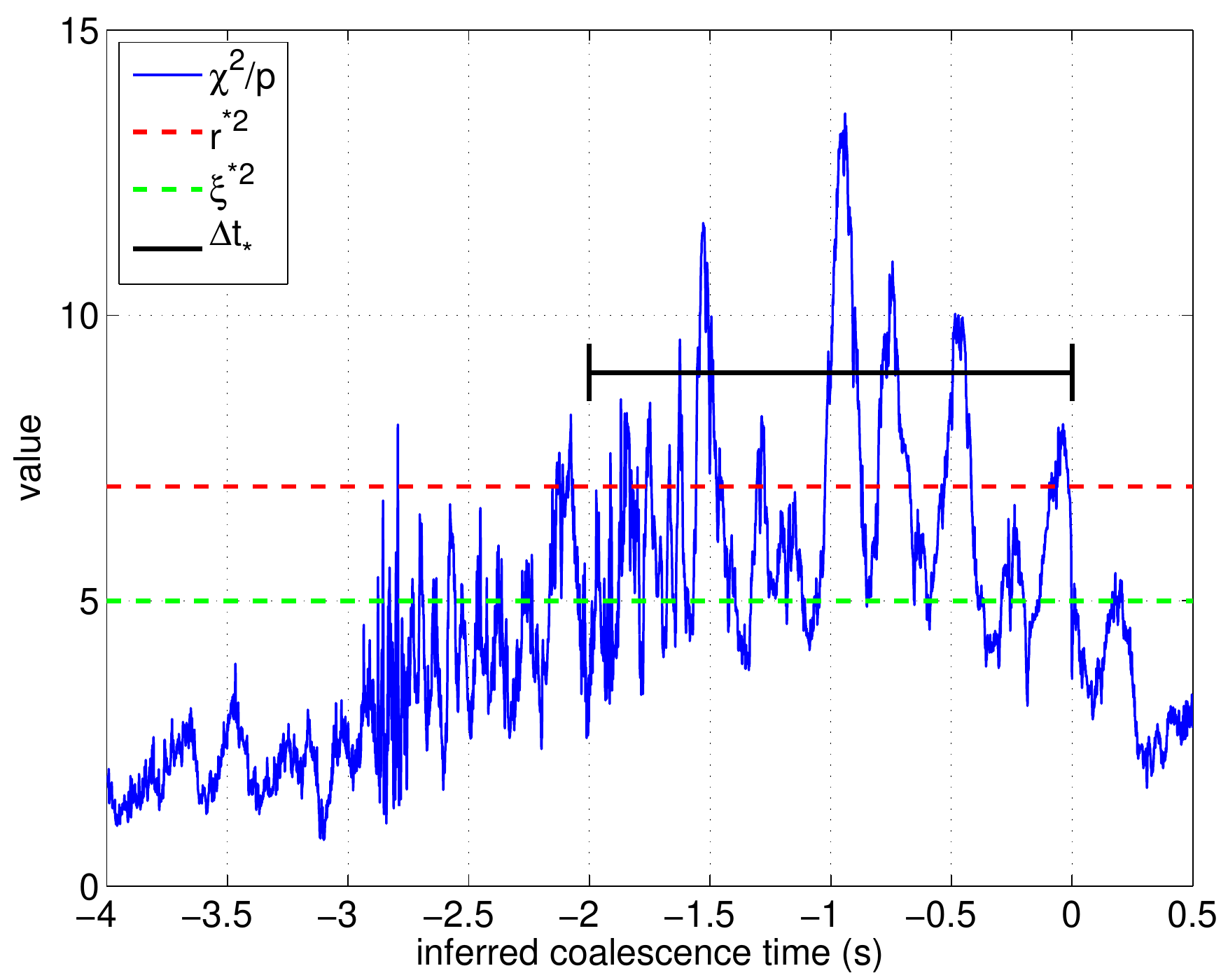} 
   \caption{\singlespace Illustration of the r$^2$ test.}
   \label{rsqtestvisual}
\end{center}
\end{figure}

\begin{figure}[!hbp|t]
\begin{center}
   \centering
   \includegraphics[height=8cm,width=10cm,angle=0]{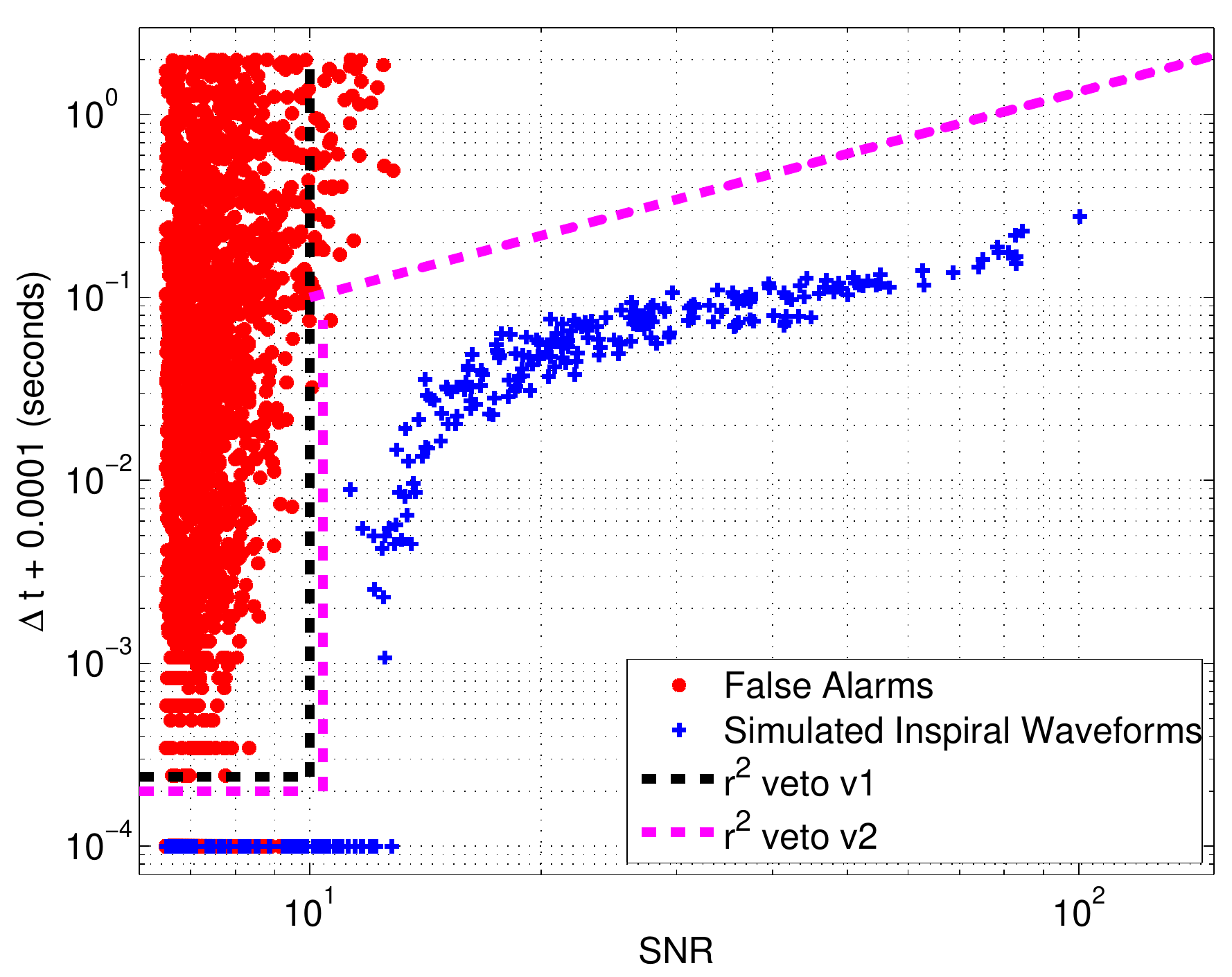} 
   \caption{\singlespace r$^2$ test example result from the S4 binary neutron star search from background triggers (time slides) and simulated inspiral injections. We chose to plot $\Delta$t plus 10 milliseconds to allow false alarms and simulated inspiral injections having $\Delta$t $\sim$0 visible in the plot. The dashed black line has the chosen parameters $\rm r_{\rm d}$ = 0.0002 and $\rho_{\rm r} = 10$. The magenta line has the chosen parameters of ${\rm r}_{\rm c}$ = 0.007 and $\rm r_{\rm p}$ = 1.1 .}
   \label{tsinjproof}
\end{center}
\end{figure}

   The r$^2$ test will require two parameters, the r$^2$ threshold, r$^{*2}$, and a window of time prior to the coalescence time to search, $\Delta$$t_{\rm *}$, which is on the order of seconds. The r$^2$ test looks up to the time before the inferred coalescence time of the trigger since the templates do not include the merger waveforms. The test will determine how much time the r$^2$ time series stays above r$^{*2}$ in $\Delta$$t_{\rm *}$, with the result denoted as $\Delta t$. A visual representation of the test is given in figure \ref{rsqtestvisual}.
An example of how the test performs for an ensemble of simulated inspiral injections and time slide triggers (false alarms) for the S4 binary neutron star search are shown in figure \ref{tsinjproof}.

In figure \ref{tsinjproof} we see two distributions. A distribution of simulated inspiral waveforms that begin to separate from the background time slide triggers (false alarms) at SNR = 10, and a distribution of false alarms that separate from the simulated waveforms for SNR $<$ 10 and $\Delta$t $>$ 0.0002 seconds. The contours plotted in the figure denote two separate methods to veto these false alarms while falsely dismissing few simulated inspiral injections. The first is a dashed black contour which includes a bottom left rectangular region that is defined by two parameters. The first is defined as a segment of constant time duration called ${\rm r}_{\rm d}$ (in the case shown in the figure, ${\rm r}_{\rm d}$ = 0.0002) which begins at the SNR threshold of the search and extends to the vertical dashed black line, which is the second parameter. The second parameter is the segment of constant SNR extending to form the long side of the rectangle called $\rho_{\rm r}$ (in the figure, $\rho_{\rm r}$ = 10). Using this black dashed contour line as a veto for this particular data set where any points within the rectangular regime are eliminated results in 35$\%$ of the false alarms being vetoed while 0.001$\%$ of the simulated inspiral injections are falsely dismissed. The use of the r$^2$ test's results as a veto is called the r$^2$ veto. The black dashed contour described in this example is called the \emph{r$^2$ veto v1}.

The second is a magenta dashed contour includes ${\rm r}_{\rm d}$, $\rho_{\rm r}$, and two additional parameters. The two additional parameters describe the power law region of the contour:

\begin{equation}
f(\rho) = {\rm r}_{\rm c}\times\rho^{{\rm r}_{\rm p}}
\end{equation}
where ${\rm r}_{\rm c}$ and ${\rm r}_{\rm p}$ are parameters chosen to create the slope. Using this magenta dashed contour line (with ${\rm r}_{\rm c}$ = 0.007, ${\rm r}_{\rm p}$  = 1.1), where any points above will be eliminated for this particular data set results in roughly 36$\%$ of the background time slide triggers being vetoed while falsely eliminating 0.001$\%$ of the simulated inspiral injections for the figure given. The magenta dashed contour described in this example is called the \emph{r$^2$ veto v2}. 

Note the four parameters can be tuned according to a given data set and inherently so the fraction of vetoed false alarms and falsely dismissed triggers will change. This will be shown in chapters \ref{chapter5} and \ref{chapter6}. The next section describes how this test could be implemented into an inspiral pipeline.

\renewcommand{\baselinestretch}{1}
\section{Implementation of the r$^2$ Test Into the Inspiral Pipeline}
\renewcommand{\baselinestretch}{2}
The r$^2$ test is implemented into the inspiral pipeline whenever the $\chi^2$ is calculated. The r$^2$ test is incorporated into the inspiral search pipeline (Sec. \ref{pipeline}) in the second matched filtering stage. The r$^2$ test pipeline is described as:
\begin{enumerate}
\item Characterize the candidate trigger by the signal to noise ratio ($\rho$), keep candidates with $\rho$ $>$ $\rho^*$.

\item  Measure the consistency of the candidate signal with the expected chirp waveform, keeping candidates with $\chi^2$$/$($p$ $+$ $\delta$$\rho^2$)  $<$ $\xi^{*2}$.

\item  Compute the time series $\chi^2$$/$$p$, count the time the trigger stays above a given time threshold, ${\rm r_{*}^2}$, in a given interval, $\Delta$t$_{\rm *}$, prior to the inferred coalescence time, $t_c$, of the trigger. Store the calculated value as $\Delta$t.
\end{enumerate}

The last step is what has been added to the inspiral pipeline. The following chapters will include examples of results from implementing the pipeline described. 

\renewcommand{\baselinestretch}{1}
\CHAPTER{The Search for Primordial Black Holes in LIGO's 3rd Science Run}
\label{chapter5}
\doublespace
A black hole composed of mass $<$ 1.0 M$_{\odot}$ is believed to be a primordial black hole (PBH) \citep{S2_5}, since it could not have been created as a product of stellar evolution. These compact objects may have formed in the early, highly compressed stages of the universe immediately following the big bang. PBH's are also theorized to be candidate Massive Astrophysical Compact Halo Objects (MACHOs), which could be some fraction of the dark matter in the Large Magellanic Cloud. A binary system composed of two PBH's will emit gravitational waves that may be detectable by LIGO \citep{Thorne87}. This chapter describes the search for primordial black hole binary systems in the third science run, including the tuning of several coincidence parameters, how the r$^2$ test was employed, and the result of the PBH binaries search from the S3 run. The results are included in \citep{LIGOS3S4all}. The results presented in this chapter are the author's while collaborating with the CBC group. 

\section{The Third LIGO Science Run}
\begin{figure}[!hbp|t]
\begin{center}
   \centering
   \includegraphics[height=10cm,width=12cm,angle=0]{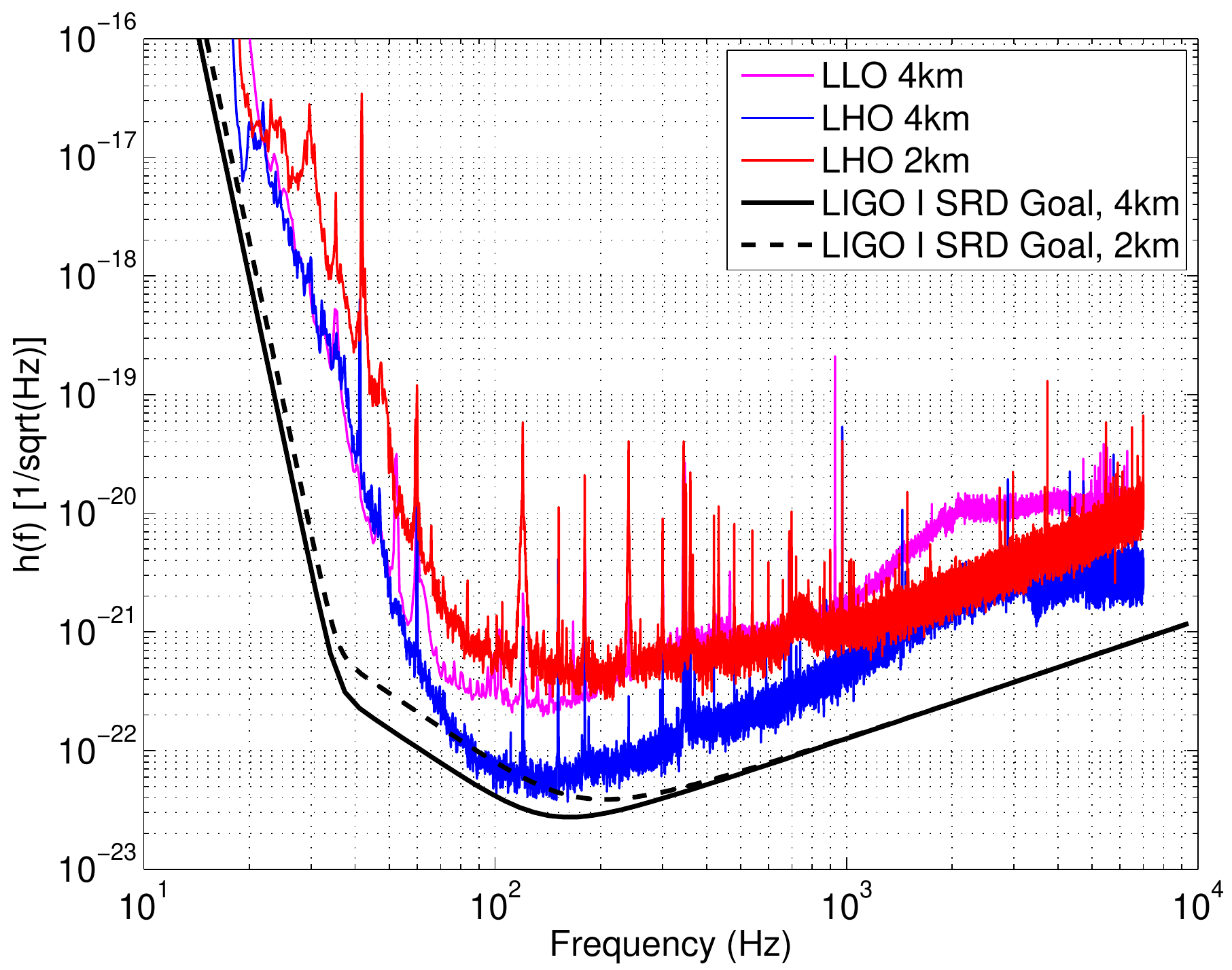} 
   \caption{\singlespace LIGO S3 Run best sensitivities for all interferometers \citep{LIGOS3strainBest}.}
   \label{LIGOS3sensitivity}
\end{center}
\end{figure}

The third LIGO science run was conducted from October 3, 2003 to January 09, 2004. All three LIGO detectors at the two observatories were in operation. The best sensitivity curve for each IFO is shown in figure \ref{LIGOS3sensitivity}. 

\begin{figure}[!hbp|t]
\begin{center}
   \centering
   \includegraphics[height=8cm,width=10cm,angle=0]{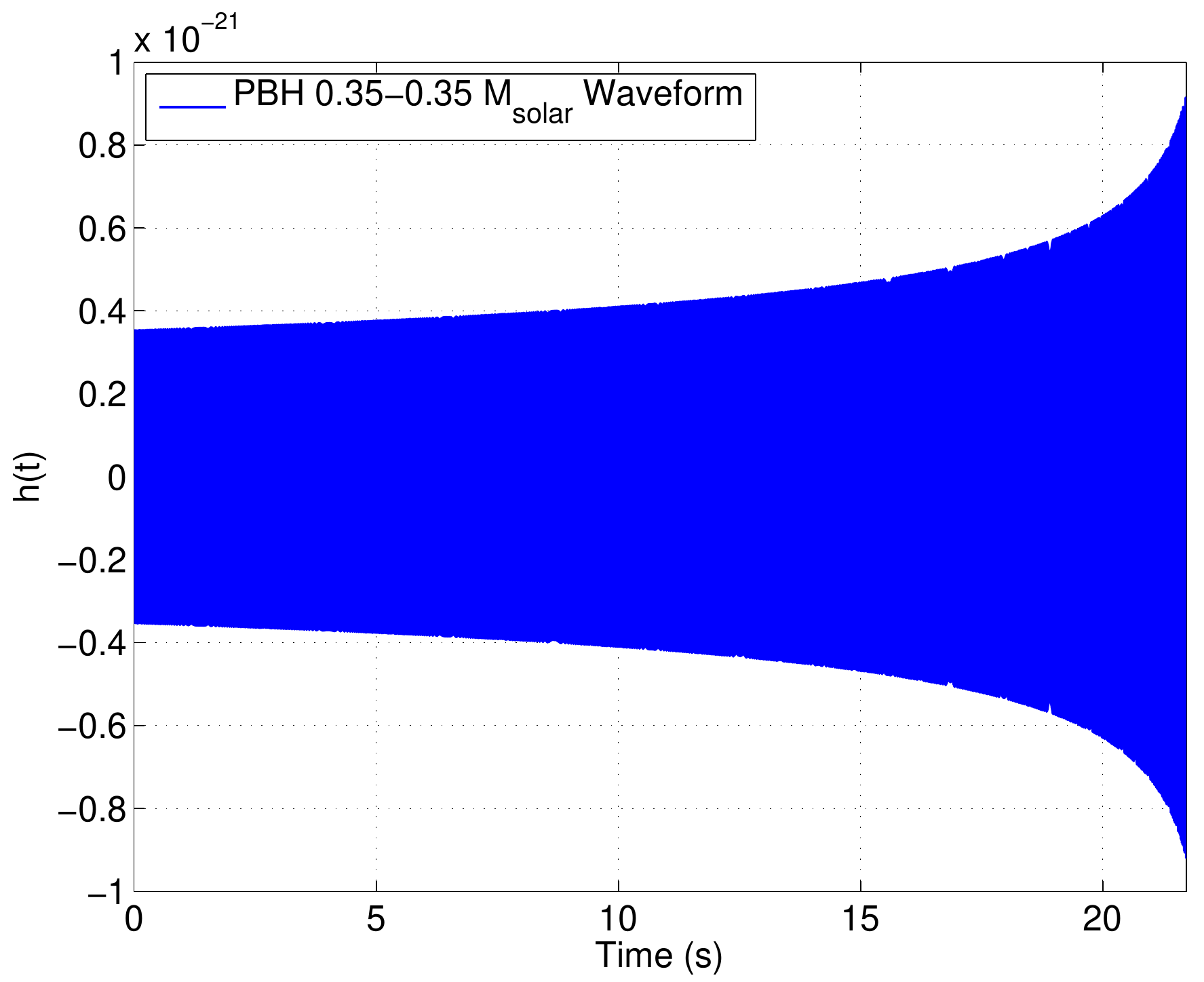} \\
   \includegraphics[height=8cm,width=10cm,angle=0]{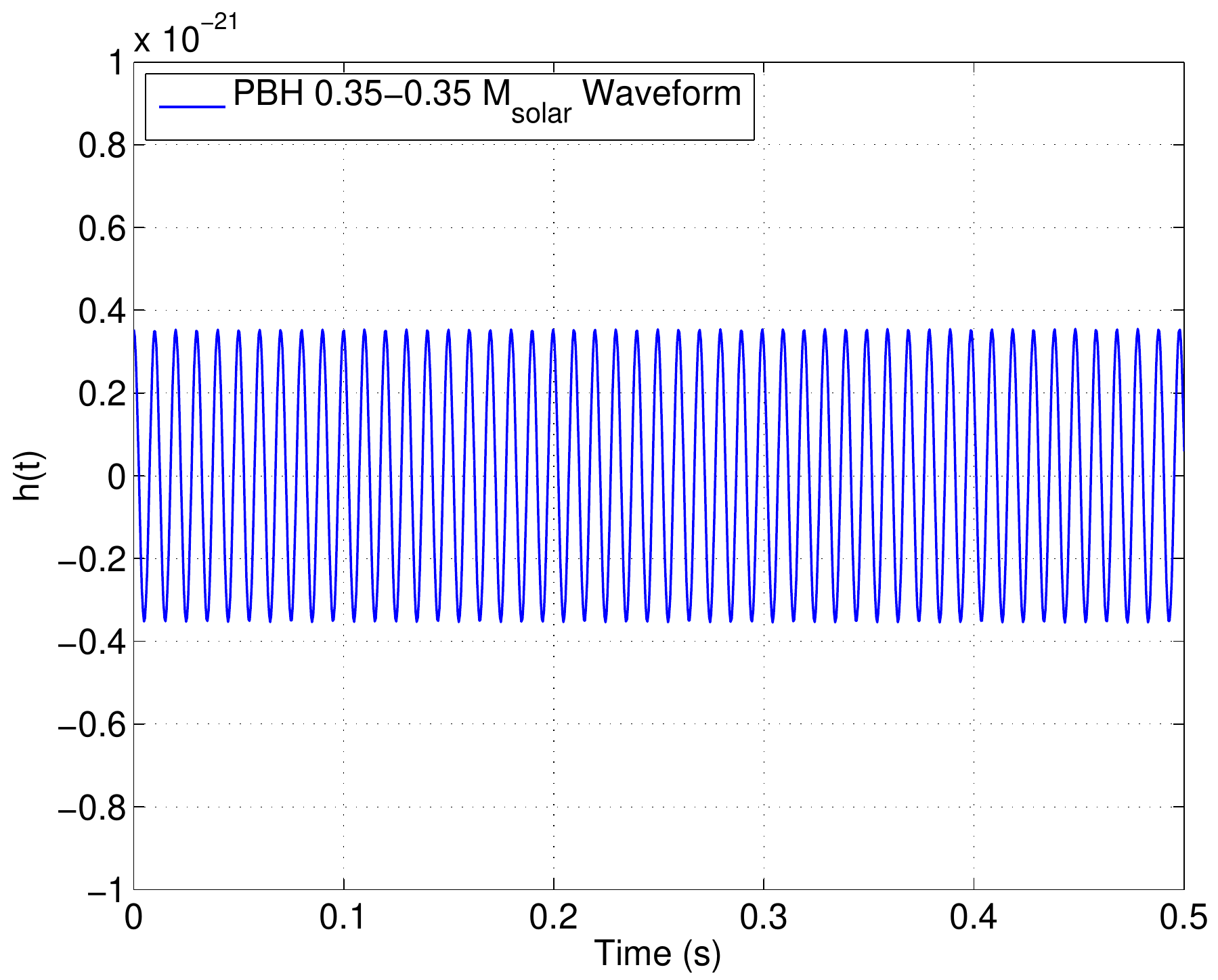} 
   \caption{\singlespace A sample PBH waveform, $f$ = 100Hz at $t$ = 0s, $f$ = 2023Hz at $t$ = 22s. A zoom of the first 0.5 seconds is shown in the bottom panel.}
   \label{S3PBH_waveform}
\end{center}
\end{figure}

Of the numerous binary systems that could be searched for gravitational wave emission, primordial black hole (PBH) binary systems were chosen to search for this thesis. The masses for the PBH search ranged from 0.35 $M_\odot$ to 1.0 $M_\odot$. We used a low frequency cutoff, $f_L$, of 100Hz. The longest waveform duration, $D_{\rm max}$, was 22.1 seconds while the average number of templates used per block of data was 4500. The range of masses for the PBH search were from 0.35 $M_\odot$ to 1.0 $M_\odot$. An example PBH template waveform is given in figure \ref{S3PBH_waveform}.

In order to search for these signals, we use a data analysis pipeline as described in section \ref{pipeline}. The first step is to determine to amount of data that will be used by collecting the times when the three interferometers were in \emph{science mode}. We also incorporate separate segment lists which include times where the quality of the data is non-optimal for each of the detectors. Once this information is collected, we calculate the times when two or more IFO's are in operation, this will represent the analyzed times to be searched and is summarized in table \ref{tab:analysedtimes}. Since second coincidence is essential to the search, a viable candidate event must have at least two detectors observing the same event. We required both of the Hanford detectors to be in operation, due to the possibility of scattered light from one of the out of lock detectors interfering with the other detector. 

\begin{table}[!htdp]
\caption{Times analyzed when at least two detectors were operating. The times
in parentheses exclude \textit{playground} times.}
\begin{center}
\begin{tabular}{lcc}
\hline
\hline
 & S3\\\hline
H1-H2-L1 times & 184 (167) hrs\\
H1-H2 times & 604 (548) hrs\\
\hline
Total times & 788 (715) hrs
\end{tabular}
\end{center}
\label{tab:analysedtimes}
\end{table}
 
\section{Tuning the Search}

In order to tune a search for inspiralling binary systems (as described in section \ref{tuning}), several parameters must be tuned. This is done is order to correctly characterize and optimize the analysis pipeline for detection. Recall that two thresholds are introduced into the pipeline at various stages in order to reduce the number of noise transients, these are the SNR and the $\chi^2$ test thresholds. We chose $\rho_{*}$ = 6.5, $\xi^{*2}$ = 10.0, with 16 $\chi^2$ bins. Recall $\rho$$(t)$ and $\xi^2$ are calculated using equations \ref{e:rho} and \ref{e:xistarsquared}. 

\begin{figure}[!hbp|t]
\begin{center}
   \centering
   \includegraphics[height=8cm,width=10cm,angle=0]{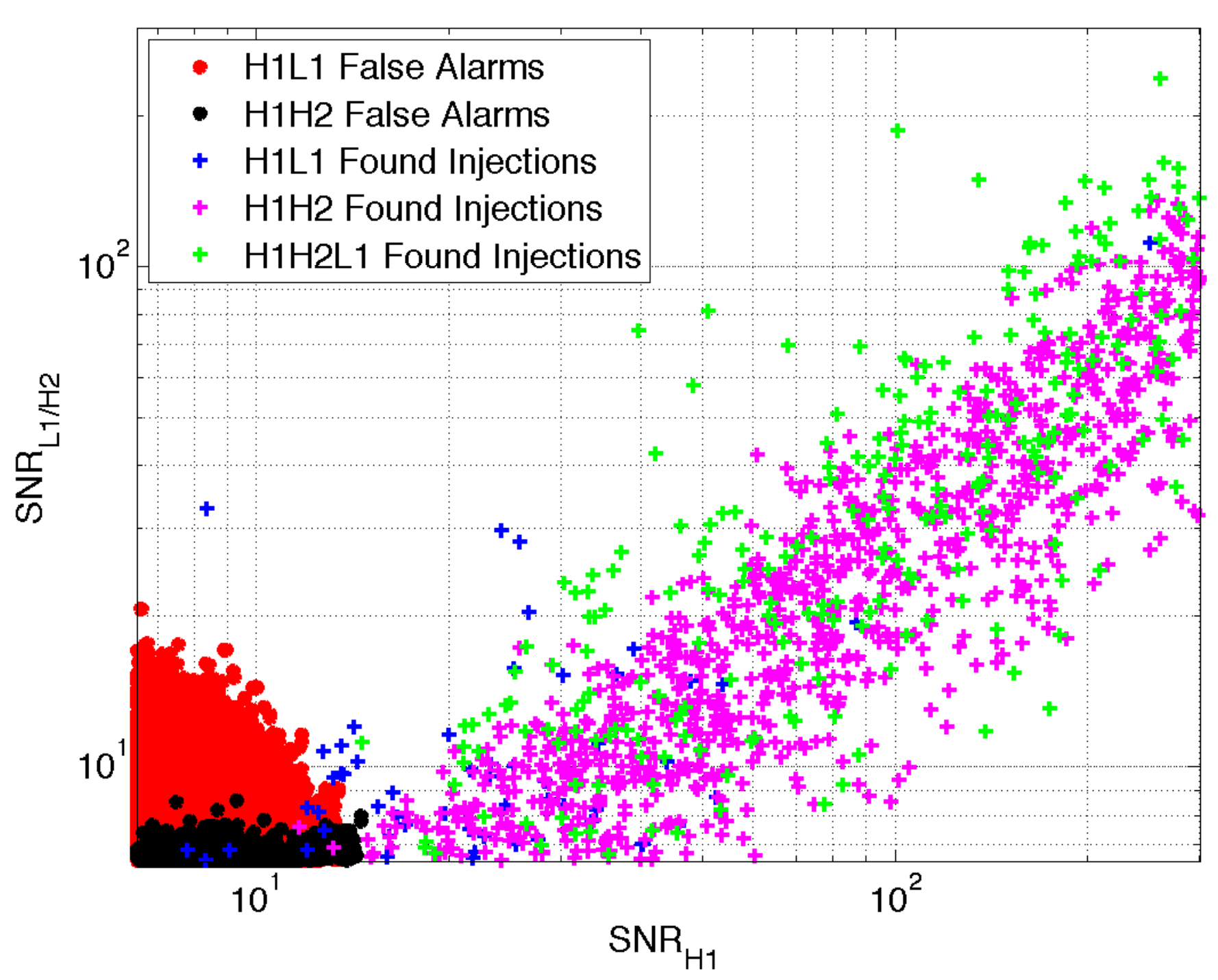} 
   \caption{\singlespace  SNR of coincident time slides (false alarms) and coincident simulated inspiral injections for the S3 PBH search.}
   \label{S3PBH_rho_allifo}
\end{center}
\end{figure}

\begin{figure}[!hbp|t]
\begin{center}
   \centering
   \includegraphics[height=8cm,width=10cm,angle=0]{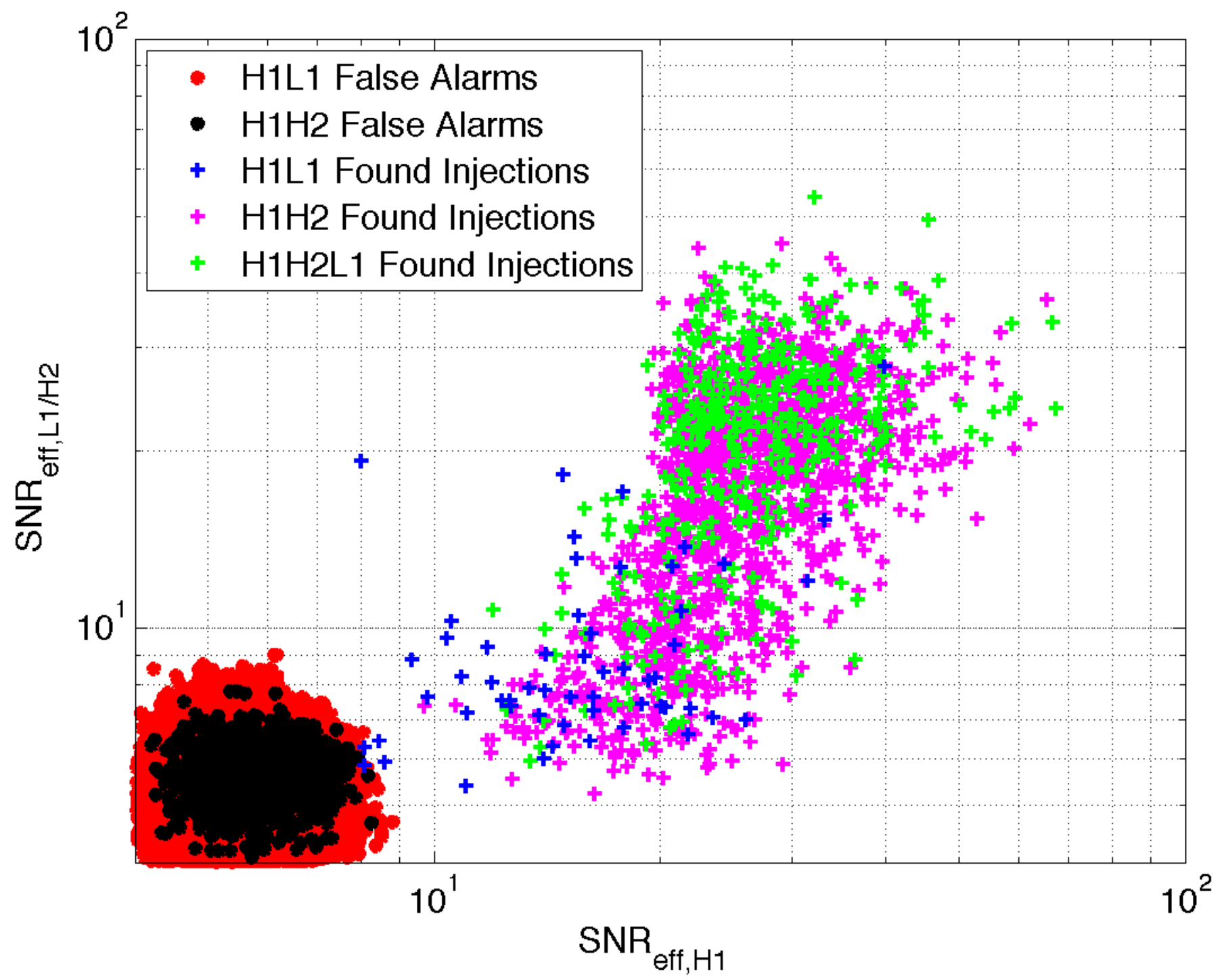} 
   \caption{\singlespace The effective SNR statistic for the S3 PBH search.}
   \label{S3PBH_rhostat_allifo}
\end{center}
\end{figure}

For the PBH binary system search, the $\chi^2$ test \cite{BAllen} provides a measure of the quality-of-fit of the signal to the template. We can also combine the SNR (equation \ref{e:rho})  and the $\chi^2$ value of the event, defined as the effective SNR, $\rho_{{\rm eff}}$, by
\begin{equation}
\label{e:rhoeff}
\rho_{\rm eff}^2 =
\frac{\rho^2}{\sqrt{\left(\frac{\chi^2}{\rm 2p-2 }\right)\left(1+\frac{\rho^2}{250}\right)}}, 
\end{equation}
where p is the number of $\chi^2$ bins used in the $\chi^2$ test; the specific value of the parameter $250$ is chosen empirically. This parameter was chosen in order to effectively separate our simulated inspiral injections with the background triggers (time slides) shown in figure \ref{S3PBH_rhostat_allifo}, since using the SNR of each coincident inspiral injections or background triggers did not effectively separate the two as shown in figure \ref{S3PBH_rho_allifo}. We expect  $\rho_{\rm eff}\sim \rho$ for real signals with relatively low SNR, and low effective SNR for noise transients with high $\chi^2$. We can also assign to each candidate in coincidence a combined SNR, $\rho_c$, defined by 
\begin{equation}
\label{eq:rhocbns}
(\rho_c)^2_{\rm PBH}=\sum_i^N \rho_{\rm eff, i}^2 
\end{equation} 
where $i$ is the detector index.

\begin{figure}[!hbp|t]
\begin{center}
   \centering
   \includegraphics[height=9cm,width=12cm,angle=0]{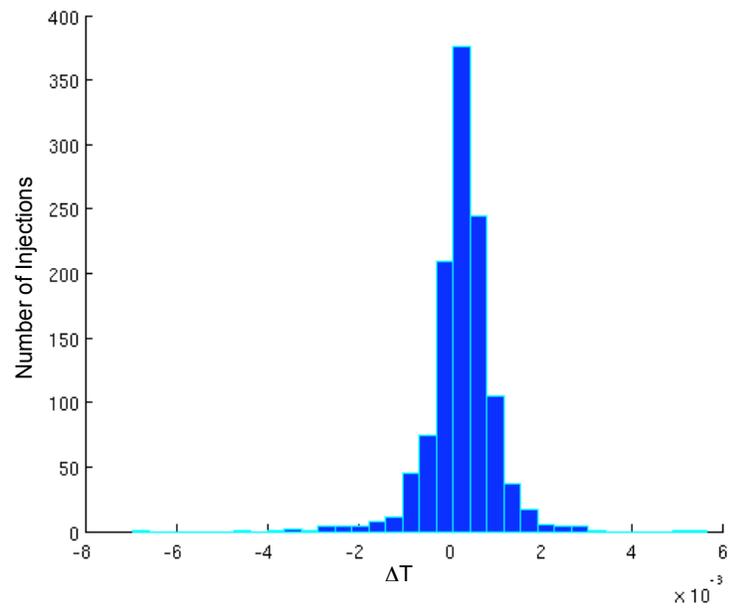} 
   \caption{\singlespace Timing difference ($\Delta$T) histogram.}
   \label{timinghisto}
\end{center}
\end{figure}

\begin{figure}[!hbp|t]
\begin{center}
   \centering
   \includegraphics[height=9cm,width=12cm,angle=0]{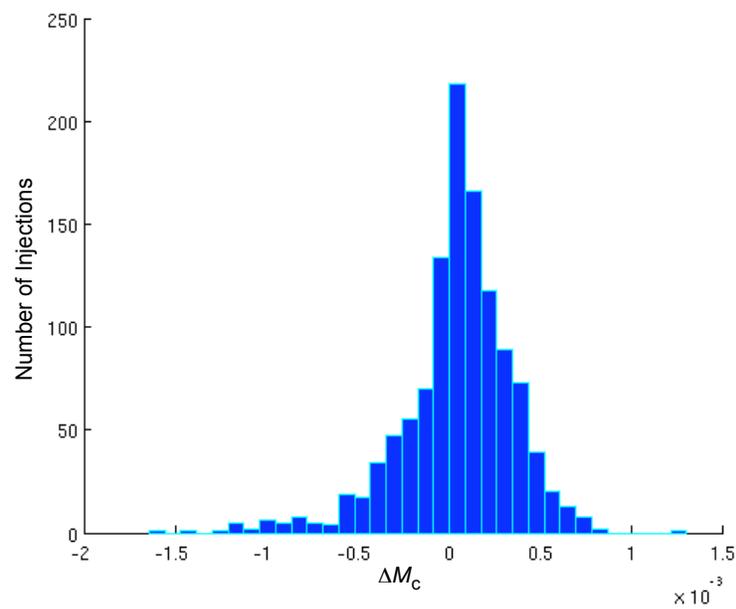} 
   \caption{\singlespace ${M}_c$ difference ($\Delta\mathcal{M}_c$) histogram.}
   \label{chirpmasshisto}
\end{center}
\end{figure}

\begin{figure}[!hbp|t]
\begin{center}
   \centering
   \includegraphics[height=9cm,width=12cm,angle=0]{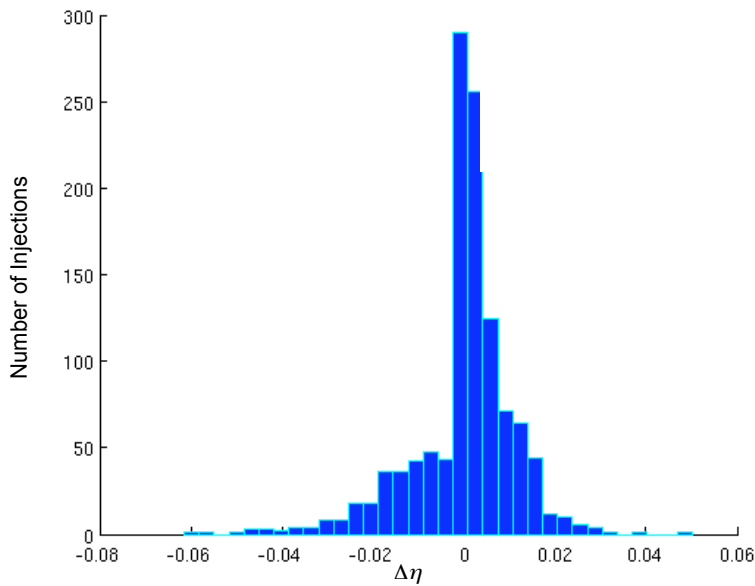} 
   \caption{\singlespace $\eta$ difference ($\Delta$$\eta$) histogram.}
   \label{etahisto}
\end{center}
\end{figure}

Several coincidence parameters are also tuned. These include the coincident time between triggers ($\Delta$T), the coincidence in the chirp mass ($\Delta\mathcal{M}_c$), and the coincidence in $\eta$ (called $\Delta$$\eta$). These parameters were tuned using results from injections. The results are plotted in figure \ref{timinghisto} for $\Delta$T, figure \ref{chirpmasshisto} for $\Delta\mathcal{M}_c$, and figure \ref{etahisto} for $\Delta$$\eta$. The chosen parameters based on these figures are given in table \ref{coincidence_params}.

\begin{table}[htdp]
\caption{\label{tab:S3PBHcoinc} Summary of the S3 PBH coincidence
windows. 
The second column gives the coincident-time windows column; we need to take
into account for time of flight between detectors (10~ms between L1 and H1/H2
detectors).}
\begin{center}
\begin{tabular}{lccc}
      \hline
      \hline
&  $\Delta$T (milliseconds)& $\Delta\mathcal{M}_c~(M_{\odot})$  & $\Delta \eta$
\\
& $4\times2$   & $0.002\times2$  & 0.06\\
\hline
\end{tabular}
\end{center}
\label{coincidence_params}
\end{table}

\begin{figure}[!hbp|t]
\begin{center}
   \centering
      \includegraphics[height=8cm,width=10cm,angle=0]{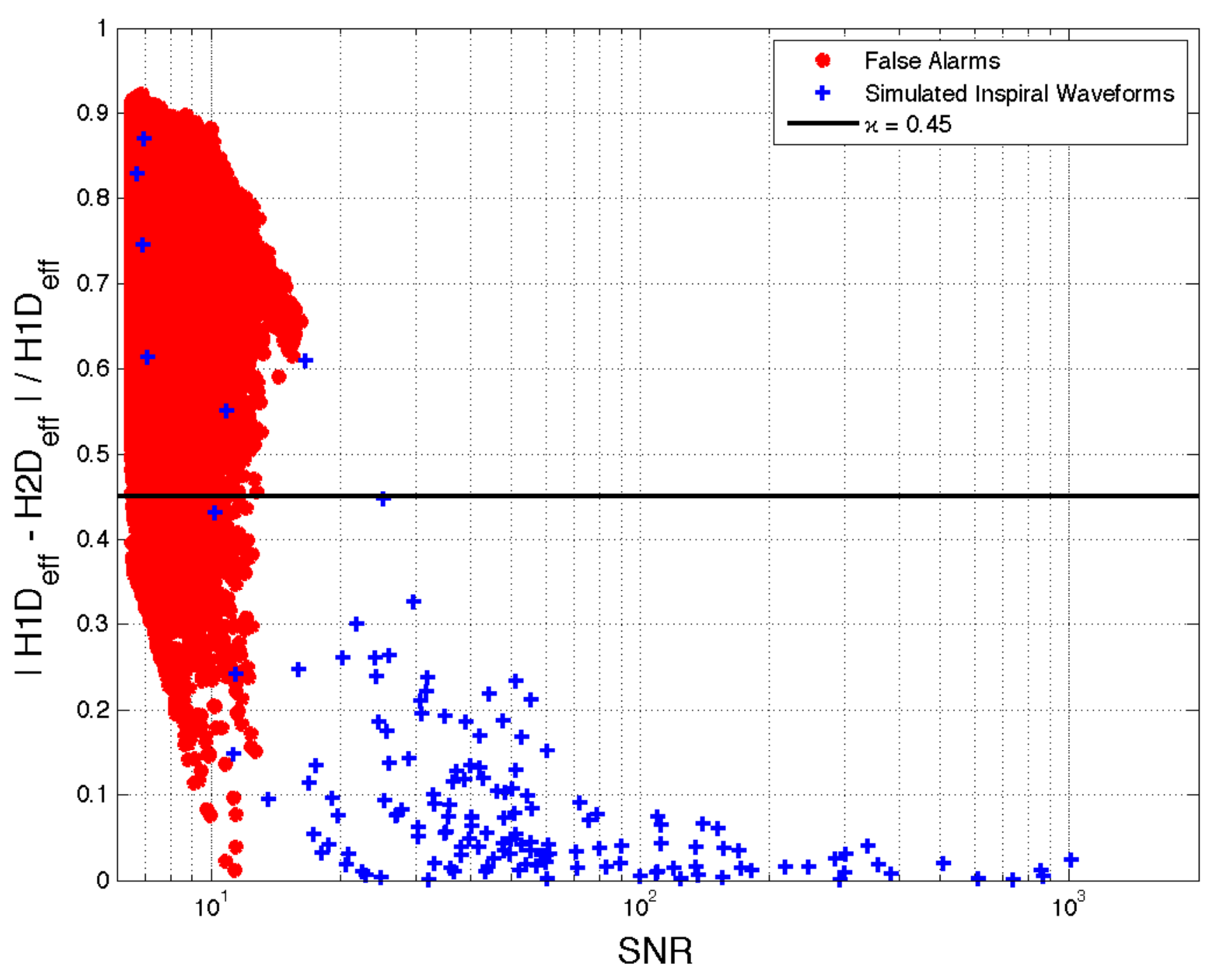} 
   \includegraphics[height=8cm,width=10cm,angle=0]{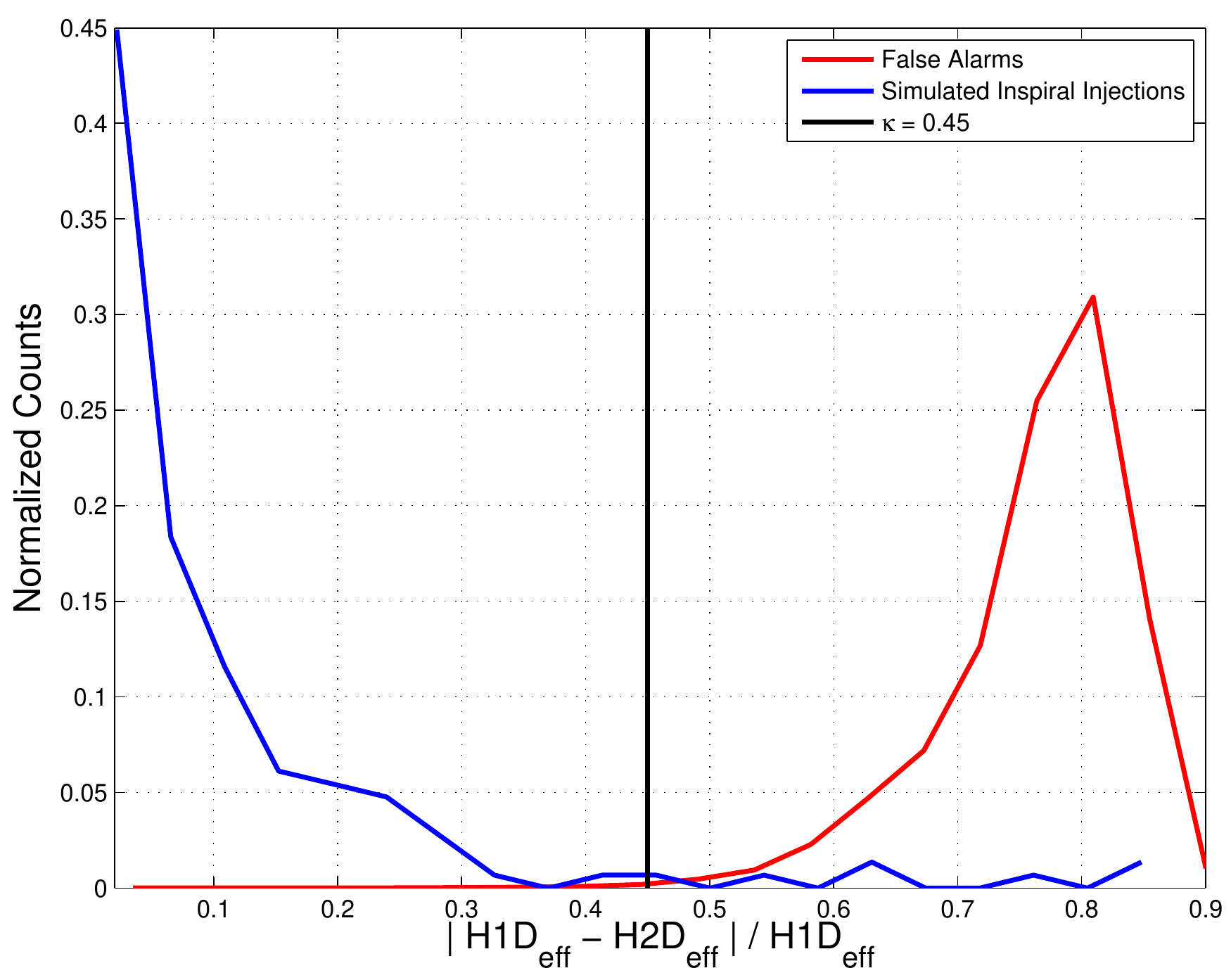} 
   \caption{\singlespace The H1H2 effective distance cut for the S3 PBH search. The bottom panel is a histogram of the H1H2 effective distance cut. The solid black line is the value of $\kappa$ chosen ($\kappa$ = 0.45).}
   \label{H1H2eff_dist_histo}
\end{center}
\end{figure}

As described in section \ref{tuning}, we can tune the error in the effective distance (equation \ref{e:kappa}) for simulations done on the H1-H2 coincident triggers by comparing the background time-slide triggers as shown in figure \ref{H1H2eff_dist_histo}, a value of $\kappa$ $=$ 0.45 was chosen.

Finally, once all the various parameters have been tuned, we run the analysis pipeline from start to end and cluster the final coincident triggers. They are clustered by locating the trigger with the largest SNR within a 22 second window, which is the length of the longest template. 

\renewcommand{\baselinestretch}{1}
\section{Results for the r$^2$ Test}
\renewcommand{\baselinestretch}{2}
The r$^2$ test as described in section \ref{rsqtest} was computed in the S3 PBH search. The parameters chosen for the test are given in table \ref{tab:rsqtest_params}.

\begin{table}[htdp]
\caption{r$^2$ Test Parameters for the S3 PBH search.}
\begin{center}
      \begin{tabular}{crrrrrr}	
      \hline
      \hline
  	\multicolumn{1}{c}{}
	& r$^{*2}$
	& $\Delta$$\rm t_{\rm *}$(seconds)
	& $\rho_{\rm r}$
	& $\rm r_{\rm d}$ (seconds)
	& $\rm r_{\rm c}$
	& $\rm r_{\rm p}$\\
	\hline	
        & 15 & 2.0 & 13 & $2\times 10^{-4}$ & - & -\\
	\hline	
      \end{tabular}
\label{tab:rsqtest_params}
\end{center}
\end{table}

\begin{figure}[!hbp|t]
\begin{center}
   \centering
   \includegraphics[height=8cm,width=10cm,angle=0]{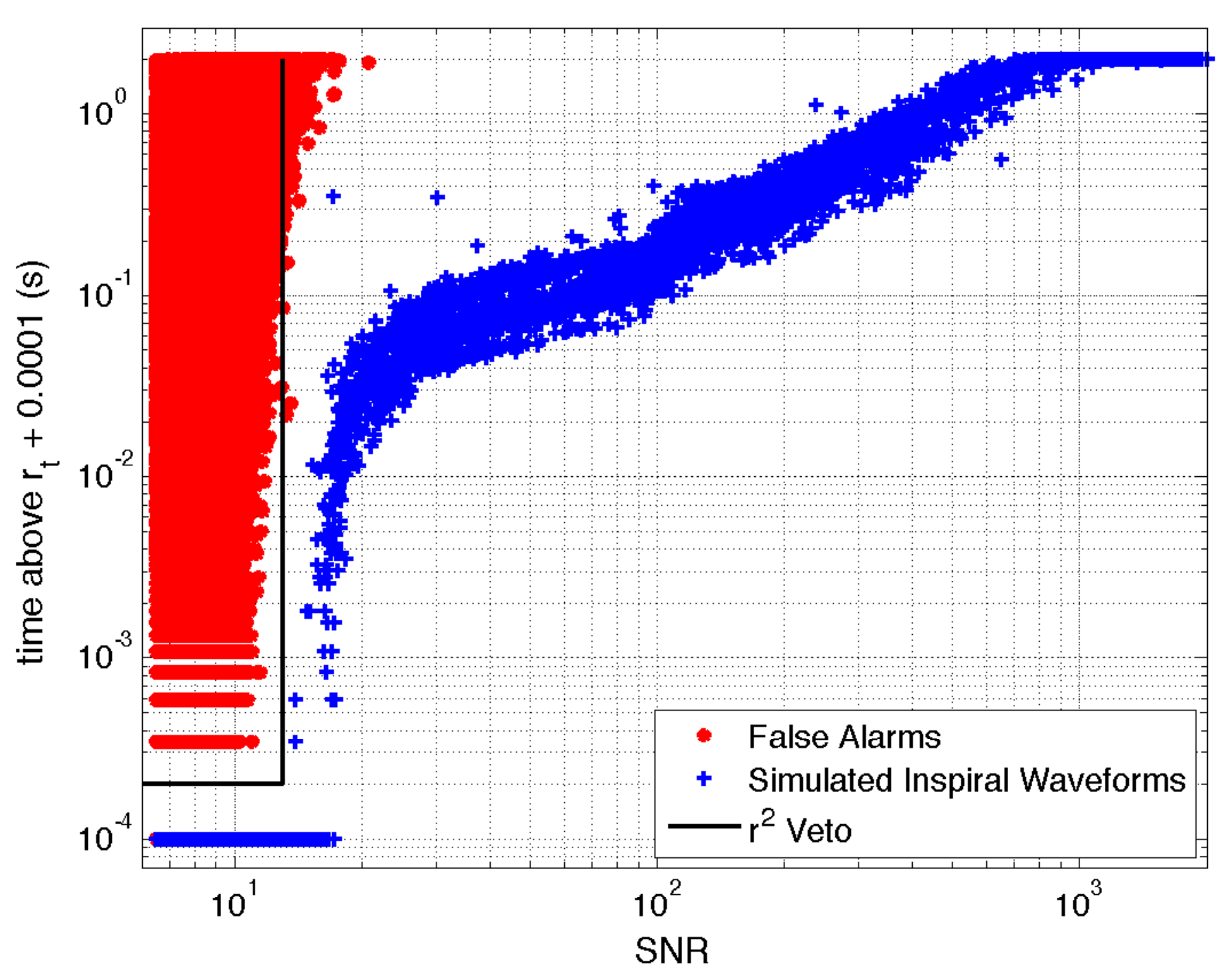} 
   \caption{\singlespace Result of the r$^2$ test  for the S3 PBH search using the parameters given in table \ref{tab:rsqtest_params}. The solid black line is the r$^2$ veto chosen, where points within it being eliminated.}
   \label{rsqtest_S3PBH}
\end{center}
\end{figure}
The results of the r$^2$ test  for the S3 PBH search are plotted in figure \ref{rsqtest_S3PBH}. We see here a clear distinction between the time slides background triggers and simulated PBH binary system  injections. The solid black line was chosen as a parameter region to veto any points within. Using the solid solid black line as a veto therefore results in eliminating 26.5$\%$ of the time slide background triggers and while falsely dismissing 0.001$\%$ of the injections. 

The result proves the power of using such a test on a binary inspiral search, where previously, coincidence between triggers and the $\chi^2$ test were the strongest vetoes available. And most importantly, it brings the possibility of finding gravitational waves in LIGO I and future detectors that much closer. In chapter \ref{chapter6}, we will see the r$^2$ test applied to several other science searches and how the veto parameters were chosen. 

\section{S3 PBH Search Result}

\begin{figure}[!hbp|t]
\begin{center}
   \centering
   \includegraphics[height=8cm,width=10cm,angle=0]{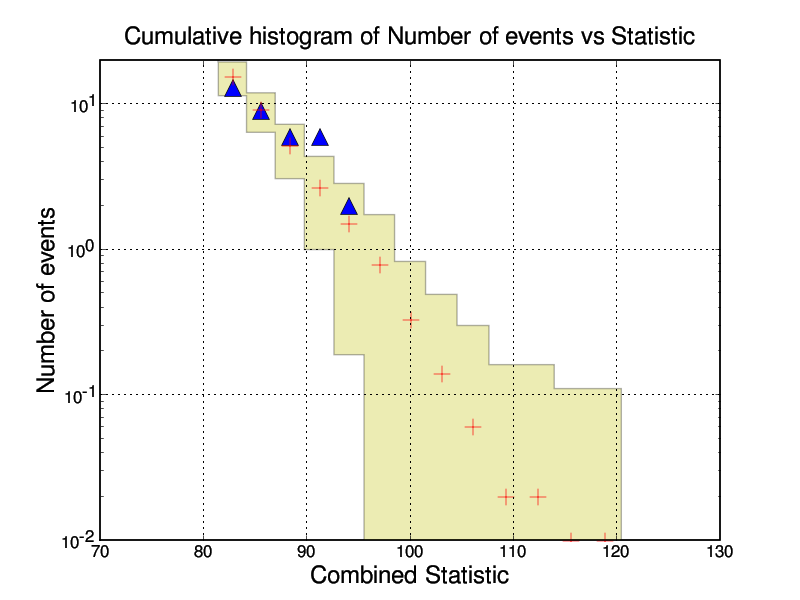} 
   \caption{\singlespace  Result of the S3 PBH search. The cumulative histogram of the combined statistic, ($\rho{_c})^{2}_{\rm PBH}$, for coincident candidates events ($\triangle$) and the estimated background from time slides (+) with 1$\sigma$ deviation ranges for the S3 PBH binaries shown (shaded region).}
   \label{foregroundvsbackground}
\end{center}
\end{figure}

A cumulative histogram of the combined statistic, ($\rho{_c})^{2}_{\rm PBH}$ as defined in equation \ref{eq:rhocbns}, of the loudest coincident triggers in the S3 PBH search is shown in figure \ref{foregroundvsbackground}. We estimate the background in the following way: 

\begin{enumerate}
\item \emph{} We shift the time of the L1 detector by 5 seconds from the true time, and the time of the H2 detector by 10 seconds from the true time. 
\item \emph{} We look for coincident triggers, which are false coincidences (thus false alarms) due to the shifted time. 
\item \emph{} We use the same pipeline (section \ref{pipeline}) used for the zero time shifted analysis, including all vetoes. 
\item \emph{} The resulting distribution of coincident triggers is considered the background of false alarms. 
\item \emph{} We repeat the procedure for 50 different time shifts, and construct a cumulative histogram of the mean number of false alarms versus combined statistic, and also calculate the standard deviation in each bin. 
\end{enumerate}

No triple coincident foreground candidate events or background events were found. We show in figure \ref{foregroundvsbackground} the foreground, or double coincidences found, and the expected background. The number of double coincidences found was consistent with the measured background. 

\renewcommand{\baselinestretch}{1}
\CHAPTER{The r$^2$ Test in Other LIGO Searches}
\label{chapter6}
\doublespace
In this chapter we revisit the r$^2$ test, as described in chapter \ref{chapter4}, applied to several other LIGO searches for gravitational waves. This includes the search for binary neutron stars (BNS) in S3/S4, the search for primordial black holes (PBH) in S4, and the search for binary black holes (BBH)  in the first three months of S5, also known as epoch 1. Each section will include a brief summary of the individual search. The searches in this chapter were done in the CBC group, where the author was a critical member on the discussions of the r$^2$ test application. 

\section{S3/S4 Binary Neutron Star Searches}
\label{s3s4_bns_rsq}

     The LIGO's third science run was conducted from October 31, 2003 to January 09, 2004 and the fourth science run ran from Feb 22, 2005 to March 24, 2005, with all all three LIGO detectors at the two observatories in operation. 
     
     The first implementation of the r$^2$ test was done on the S3 binary neutron star search. The BNS searches \cite{LIGOS3S4all} used post-Newtonian templates in the range 1.0 $M_\odot < m_1,m_2 < 3.0 M_\odot$, which for the S3/S4 searches are summarized in table \ref{tab:params_S3S4BNS}, where $N_{\rm b}$ is the number of templates in the bank and $D_{\rm max}$ is the length of the longest template. The coincidence parameters are listed in table \ref{tab:params_coinc_S3S4BNS}. 
    
\begin{table}[!htdp]
\caption{\singlespace Target Sources of the S3/S4 BNS Search. }
\begin{center}
      \begin{tabular}{crrrrrr}	
      \hline
      \hline
  	\multicolumn{1}{c}{}
	& $m_{\rm min}(M_\odot)$
	& $m_{\rm max}(M_\odot)$	
	& $f_L$(Hz)
	& $N_{\rm b}$ & $D_{\rm max}$(s)\\\hline
	S3 BNS & 1.0 & 3.0 &70 & 2000&10.0\\
	S4 BNS & 1.0 & 3.0 &40 & 3500&44.4\\
	\hline	
      \end{tabular}
\label{tab:params_S3S4BNS}
\end{center}
\end{table}

\begin{table}[!htdp]
\caption{Summary of the S3/S4 BNS Coincidence Windows.}
\begin{center}
\begin{tabular}{lccc}
      \hline
      \hline
&  $\Delta$T (milliseconds) & $\Delta\mathcal{M}_c~(M_{\odot})$ & $\Delta$$\eta$ \\\hline
S3/S4 BNS & $5\times2$   & $0.01 \times2$  & 0.10\\
\hline	
\end{tabular}
\label{tab:params_coinc_S3S4BNS}
\end{center}
\end{table}

\begin{figure}[!hbp|t]
\begin{center}
   \centering
   \includegraphics[height=8cm,width=10cm,angle=0]{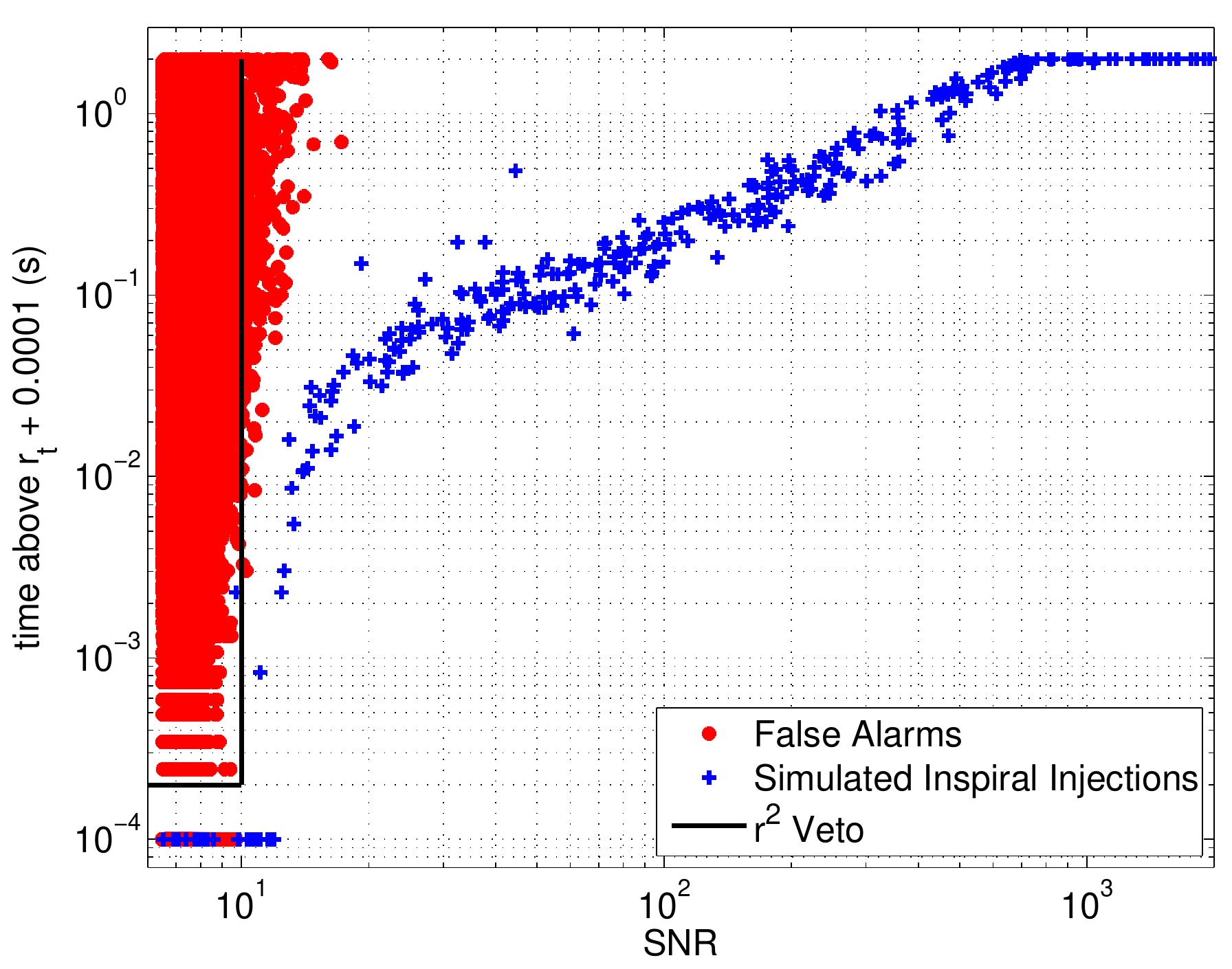} 
   \caption{\singlespace Result of the r$^2$ test  for the S3 BNS search using the parameters given in table \ref{tab:rsqtest_params_S3S4BNS}. The solid black line is the r$^2$ veto chosen, where points within it being eliminated.}
   \label{fig:rsqtest_params_S3BNS}
\end{center}
\end{figure}

\begin{figure}[!hbp|t]
\begin{center}
   \centering
   \includegraphics[height=8cm,width=10cm,angle=0]{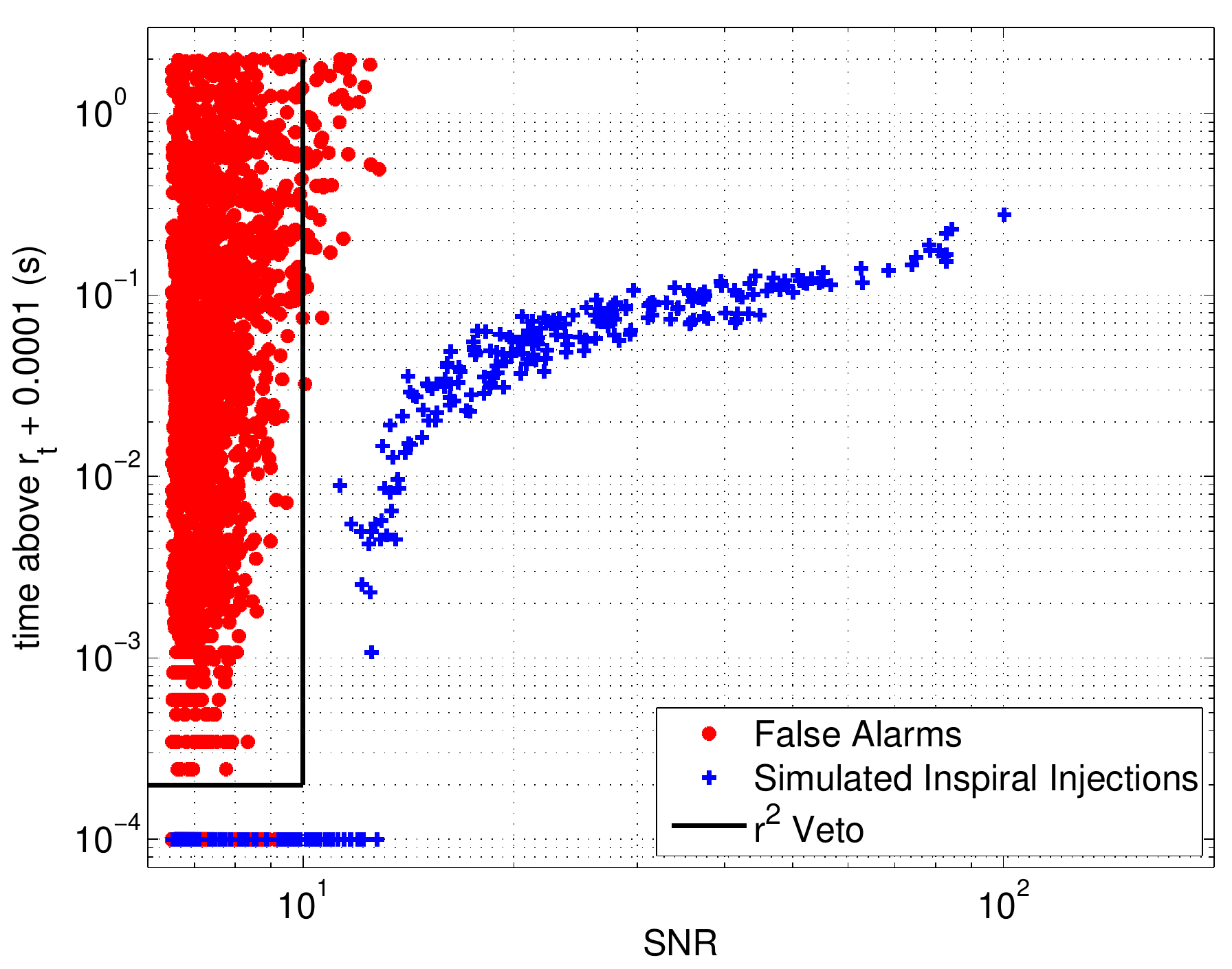} 
   \caption{\singlespace Result of the r$^2$ test  for the S4 BNS search using the parameters given in table \ref{tab:rsqtest_params_S3S4BNS}. The solid black line is the r$^2$ veto chosen, where points within it being eliminated.}
   \label{fig:rsqtest_params_S4BNS}
\end{center}
\end{figure}

    The r$^2$ test  parameters in the S3/S4 BNS search are given in table \ref{tab:rsqtest_params_S3S4BNS}. The results of the test are given in figures \ref{fig:rsqtest_params_S3BNS} and \ref{fig:rsqtest_params_S4BNS}, where the black line denotes the r$^2$ veto chosen. The use of the r$^2$ veto significantly lowers the background, vetoing about 43$\%$ of the false alarms for S3 and 35$\%$ for S4, even after the $\chi^2$ veto and other vetoes. The veto falsely dismisses only 0.001$\%$ of the simulated inspiral injections for both runs.

\begin{table}[!htdp]
\caption{r$^2$ Test Parameters for the S3/S4 BNS Search.}
\begin{center}
      \begin{tabular}{crrrrrr}	
      \hline
      \hline
  	\multicolumn{1}{c}{}
	& r$^{*2}$
	& $\Delta$$\rm t_{\rm *}$ (seconds)
	& $\rho_{\rm r}$
	& $\rm r_{\rm d}$ (seconds)
	& $\rm r_{\rm c}$
	& $\rm r_{\rm p}$\\
	\hline
        S3/S4 BNS & 10 & 2.0 & 10 & $2\times 10^{-4}$ & - & -\\
	\hline	
      \end{tabular}
\label{tab:rsqtest_params_S3S4BNS}
\end{center}
\end{table}

\begin{table}[!htdp]
\caption{r$^2$ Test Results for S3/S4 BNS Searches}
\begin{center}
      \begin{tabular}{crrrrrr}	
      \hline
      \hline
  	\multicolumn{1}{c}{}
	& Falsely Dismissed Injections (S3,S4)
	& Vetoed False Alarms (S3,S4)\\
	\hline
       \% & 0.001, 0.001  & 43.0, 35.0 \\
	\hline	
      \end{tabular}
\label{tab:rsqtest_results_S3S4BNS}
\end{center}
\end{table}

\section{S4 Primordial Black Hole Search}

 The S4 PBH searches \cite{LIGOS3S4all} use post-Newtonian templates in the range 0.35 $M_\odot < m_1,m_2 < 1.0 M_\odot$, which is summarized in table \ref{tab:params_S4PBH}. The coincidence parameters chosen are included in table \ref{tab:params_coinc_S4PBH}.
     
\begin{table}[!htdp]
\caption{Target Sources of the S4 PBH search.}
\begin{center}
      \begin{tabular}{crrrrrr}	
      \hline
      \hline
  	\multicolumn{1}{c}{}
	& $m_{\rm min}(M_\odot)$
	& $m_{\rm max}(M_\odot)$	
	& $f_L$(Hz)
	& $N_{\rm b}$ & $D_{\rm max}$(s)\\\hline
	& 0.35 & 1.0 & 100 & 4500&22.1\\\hline	
      \end{tabular}
\label{tab:params_S4PBH}
\end{center}
\end{table}

\begin{table}[!htdp]
\caption{Summary of the S4 PBH coincidence windows.}
\label{tab:params_coinc_S4PBH}
\begin{center}
\begin{tabular}{lccc}
      \hline
      \hline
&  $\Delta$T (milliseconds) & $\Delta\mathcal{M}_c~(M_{\odot})$   & $\Delta \eta$ 
\\
		& $4\times2$   & $0.002\times2$ & 0.06 \\
\hline	
\end{tabular}
\end{center}
\end{table}

\begin{figure}[!hbp|t]
\begin{center}
   \centering
   \includegraphics[height=8cm,width=10cm,angle=0]{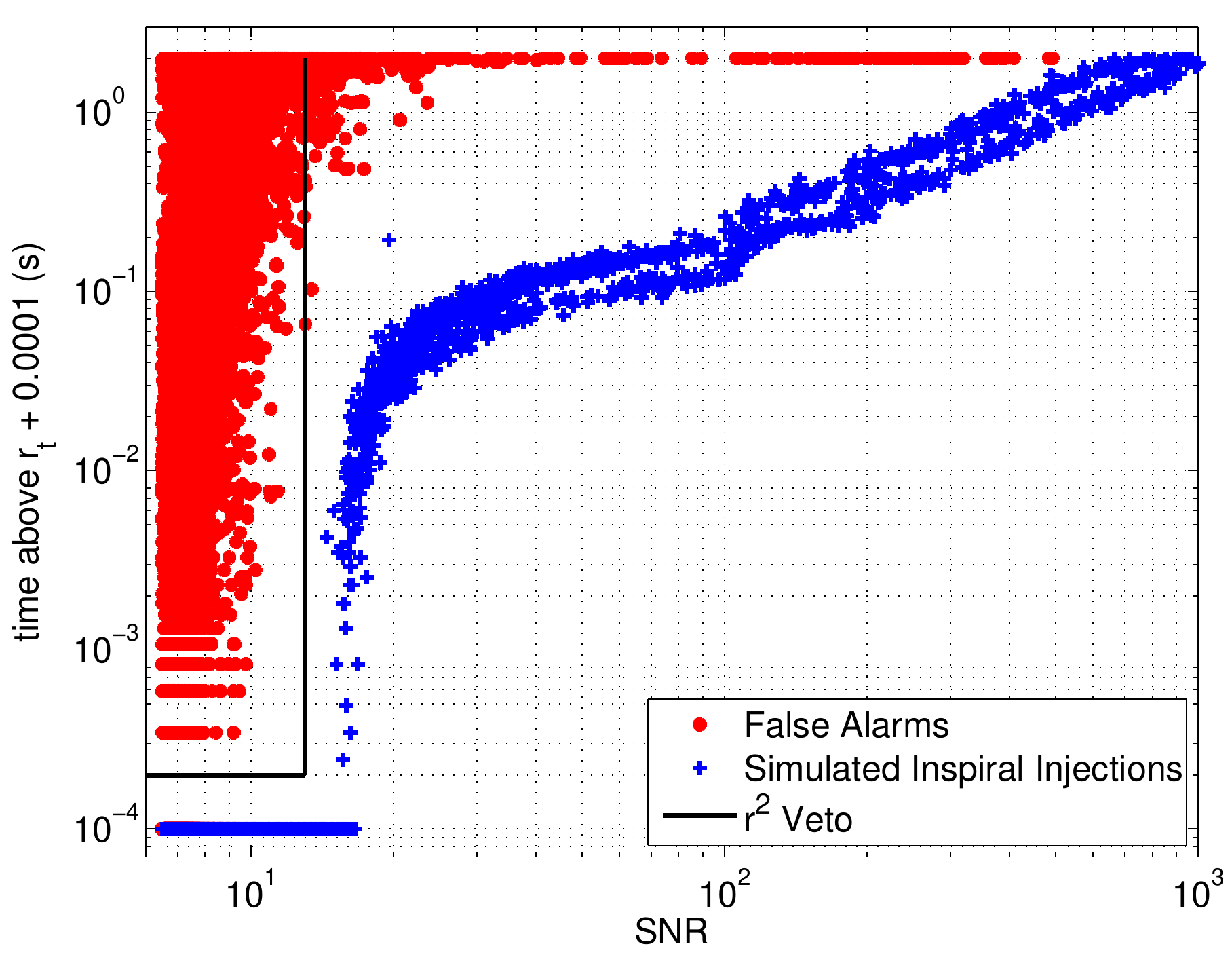} 
   \caption{\singlespace Result of the r$^2$ test  for the S4 PBH search using the parameters given in table \ref{tab:rsqtest_params_S4PBH}. The solid black line is the r$^2$ veto chosen, where points within it being eliminated.}
      \label{fig:rsqtest_params_S4PBH}
\end{center}
\end{figure}

    The r$^2$ test  was done on the S4 primordial black hole search \cite{LIGOS3S4all} . The parameters are given in table \ref{tab:rsqtest_params_S4PBH}. The results of the test are given in figure \ref{fig:rsqtest_params_S4PBH}, where using the black line is the r$^2$ veto chosen and results in gives a significant lowering of the background on the order of 35$\%$ after the $\chi^2$ veto and other vetoes, while falsely dismissing 0.001$\%$ of the simulated inspiral injections. 
    
\begin{table}[!htdp]
\caption{r$^2$ Test Parameters for the S4 PBH Search.}
\begin{center}
      \begin{tabular}{crrrrrr}	
      \hline
      \hline
  	\multicolumn{1}{c}{}
	& r$^{*2}$
	& $\Delta$$\rm t_{\rm *}$ (seconds)
	&$\rho_{\rm r}$
	& $\rm r_{\rm d}$ (seconds)
	& $\rm r_{\rm c}$
	& $\rm r_{\rm p}$\\
	\hline
        S4 & 15 & 2.0 & 13 & $2\times 10^{-4}$ & - & -\\
	\hline	
      \end{tabular}
\label{tab:rsqtest_params_S4PBH}
\end{center}
\end{table}

\begin{table}[!htdp]
\caption{r$^2$ test  Results for S4 PBH Search}
\begin{center}
      \begin{tabular}{crrrrrr}	
      \hline
      \hline
  	\multicolumn{1}{c}{}
	& Falsely Dismissed Injections
	& Vetoed False Alarms\\
	\hline
       \% & 0.001 & 35.0\\
	\hline	
      \end{tabular}
\label{tab:rsqtest_results_S4PBH}
\end{center}
\end{table}

\section{S5 Binary Black Hole Search (Epoch 1)}

\begin{figure}[!hbp|t]
\begin{center}
   \centering
   \includegraphics[height=8cm,width=10cm,angle=0]{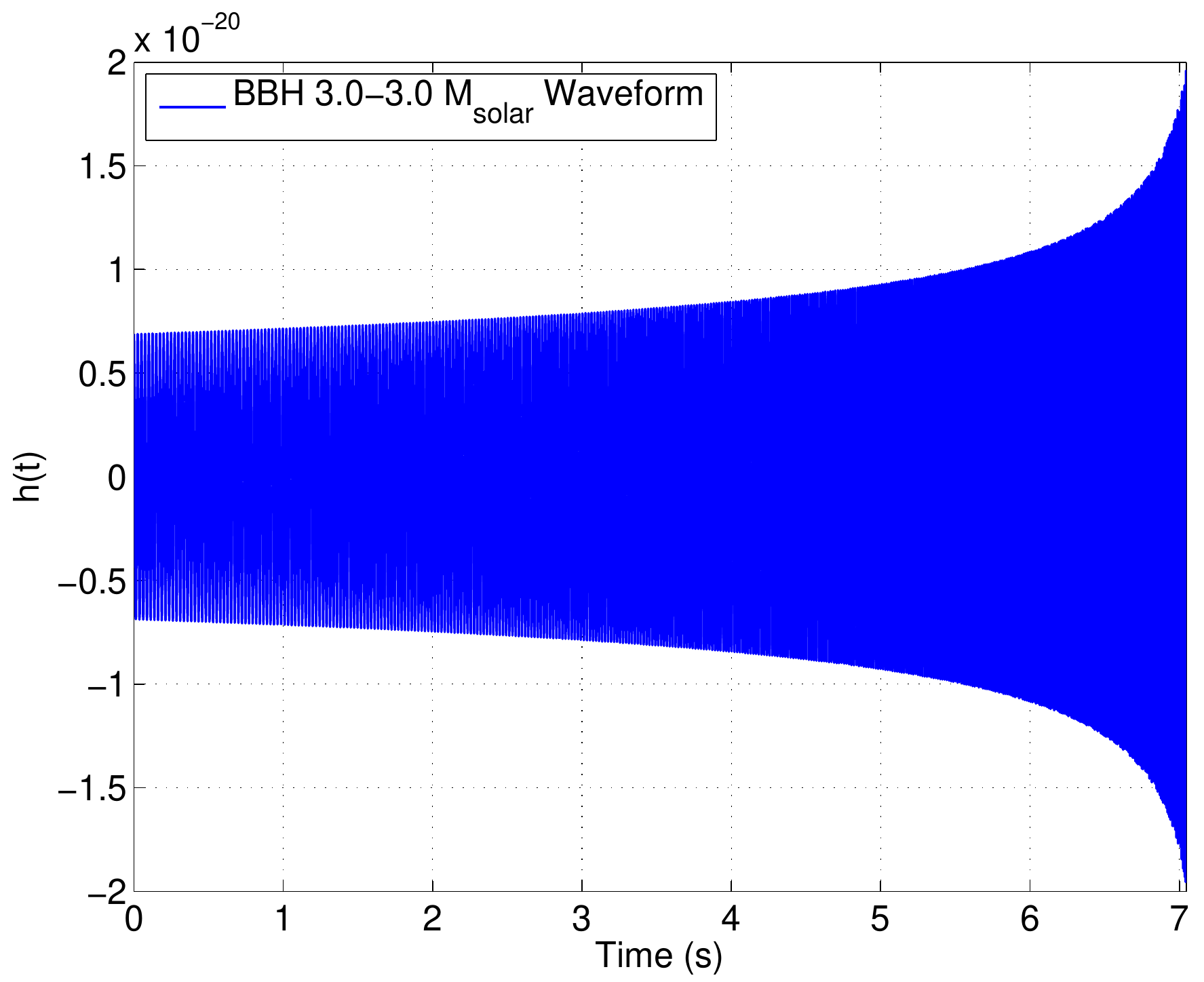} \\
   \includegraphics[height=8cm,width=10cm,angle=0]{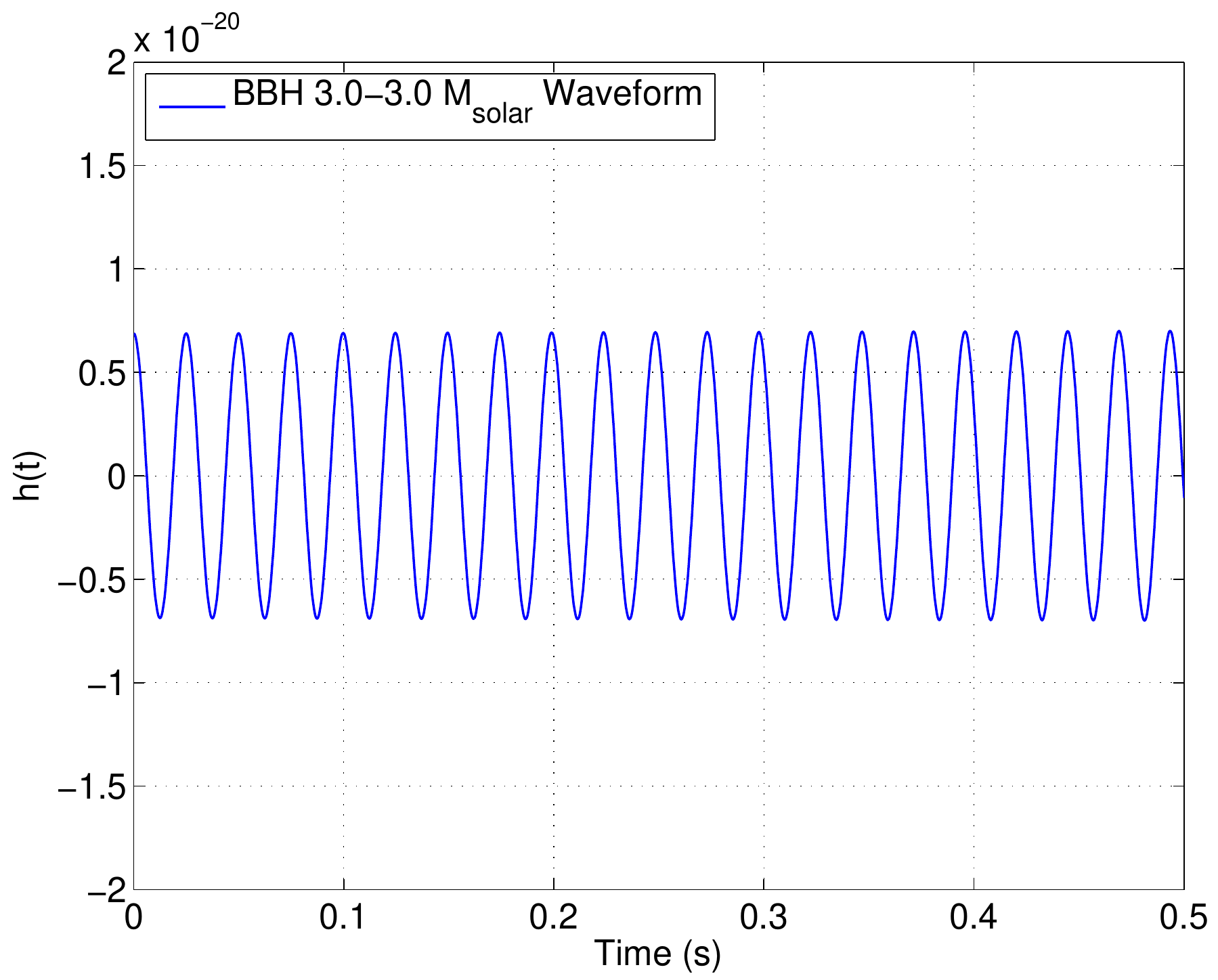} 
   \caption{\singlespace A sample BBH waveform, $f$ = 40Hz at $t$ = 0s, $f$ = 701Hz at $t$ = 7s. A zoom of the first 0.5 seconds is shown in the bottom panel.}
   \label{S5BBH_waveform}
\end{center}
\end{figure}

S5 started on November 4, 2005. We analyze the ``first epoch'', from November 4, 2005 to February 6, 2006, with all all three LIGO detectors at the two observatories were in operation. This search targeted black hole binaries of component mass in the range 3.0 $M_\odot < m_1,m_2 < 30.0 M_\odot$, with of total mass no higher than 35.0 $M_{\odot}$. For the S5 epoch 1 search, we decided to use standard post-Newtonian templates with a $\chi^2$  test. A sample waveform is shown in figure \ref{S5BBH_waveform}. BBH simulated inspiral injections were done from 4 different waveform families which included Pade-T1, Taylor-T1, Taylor-T3, and Effective-One-Body \citep{BuonannoChenVallisneri:2003a}.

\begin{figure}[!hbp|t]
\begin{center}
   \centering
   \includegraphics[height=8cm,width=10cm,angle=0]{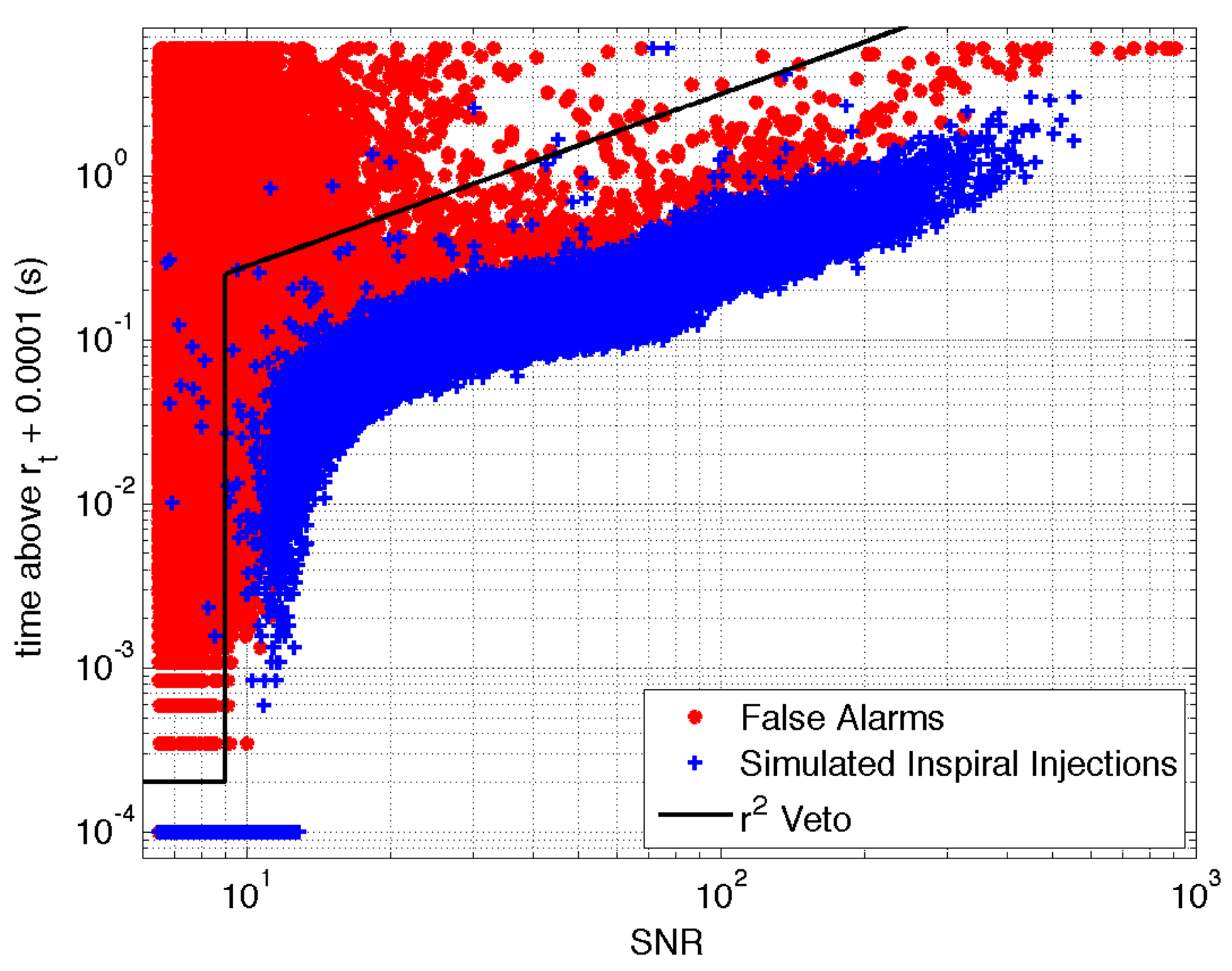} 
   \caption{\singlespace Result of the r$^2$ test  for the S5 BBH search using the values given in table \ref{tab:rsqtest_params_S5BBH}. The solid black line is the r$^2$ veto chosen, where points within it being eliminated.}
\label{fig:rsqtest_params_S5BBH}
\end{center}
\end{figure}

    Since the $\chi^2$ test was used, we could therefore incorporate the r$^2$ test  as well. The parameters for the r$^2$ test is given in table \ref{tab:rsqtest_params_S5BBH}. The results of the test are given in figure \ref{fig:rsqtest_params_S5BBH}. Note this was the first time we decided to use a non-zero value of $\rm r_{\rm p}$. In figure \ref{fig:rsqtest_params_S5BBH} we see an overlap between the false alarm triggers and simulated inspiral injections, possibly due to the shorter length of the BBH templates. This results in not being able to veto most false alarms beyond a given SNR as we saw in section \ref{s3s4_bns_rsq}, but does significantly lower the background on the order of 19$\%$.  
        
\begin{table}[!htdp]
\caption{The target sources of the S5 BBH search.}
\begin{center}
      \begin{tabular}{crrrrrr}	
      \hline
      \hline
  	\multicolumn{1}{c}{}
& $m_{\rm min}(M_\odot)$
	& $m_{\rm max}(M_\odot)$	
	& $f_L$(Hz)
	& $N_{\rm b}$ & $D_{\rm max}$(s)\\\hline
	 & 3.0 & 25.0 & 40 & 300 & 7.0\\\hline	
      \end{tabular}
\label{tab:params_S5BBH}
\end{center}
\end{table}

\begin{table}[!htdp]
\caption{\label{tab:coinc} Summary of the S5 BBH coincidence
windows.}
\begin{center}
\begin{tabular}{lcc}
      \hline
      \hline
&  $\Delta$T (milliseconds) & $\Delta\mathcal{M}_c~(M_{\odot})$ \\
& $20\times2$   & $5.0 \times2$  \\
\hline	
\end{tabular}
\end{center}
\end{table}

\begin{table}[!htdp]
\caption{r$^2$ Test Parameters for the S5 BBH search.}
\begin{center}
      \begin{tabular}{crrrrrr}	
      \hline
      \hline
  	\multicolumn{1}{c}{}
	& r$^{*2}$
	& $\Delta$$\rm t_{\rm *}$ (seconds)
	& $\rho_{\rm r}$
	& $\rm r_{\rm d}$ (seconds)
	& $\rm r_{\rm c}$
	& $\rm r_{\rm p}$\\
	\hline
        S5 & 10 & 6.0 & 9 & $2\times 10^{-4}$ & 0.025 & 1.05\\
	\hline	
      \end{tabular}
\label{tab:rsqtest_params_S5BBH}
\end{center}
\end{table}

\begin{table}[htdp]
\caption{r$^2$ test  Results for S5 BBH epoch 1}
\begin{center}
      \begin{tabular}{crrrrrr}	
      \hline
      \hline
  	\multicolumn{1}{c}{}
	& Falsely Dismissed Injections
	& Vetoed False Alarms\\
	\hline
       \% & 0.12 & 19.1\\
	\hline	
      \end{tabular}
\label{tab:rsqtest_results_S5BBH}
\end{center}
\end{table}


\CHAPTER{Conclusions}
\label{chapter7}
\doublespace

We have devised a new test, that we call the r$^2$ test, that greatly reduces the number of false alarms in gravitational wave searches of coalescing binary systems. We have shown the actual efficiency of this test using data from the LIGO interferometers from several science runs. The reduction in the rate of false alarms ranged for example from 43$\%$ for the search for binary neutron star systems in the third science run, to 19$\%$ in the search for the coalescence of binary black hole systems in the first three months of the fifth science run, while achieving a low false dismissal rate of simulations ranging from 0.001$\%$ to 0.12$\%$, respectively. The $r^2$ test will be incorporated into future LIGO searches such as the searches for binary systems in the fifth science run. A search for primordial black hole binary systems (where each object has less than 1M$_\odot$) in LIGO's Third Science Run (S3) was conducted where results from the number of double coincidences found was consistent with the measured background.

\singlespace

\singlespace
\TOPPER{Vita}
\doublespace
Andr\'es Rodr\'iguez was born on January 31, 1979 in Miami, Florida.  He went to high school at Belen Jesuit Preparatory School, where a junior year course in introductory chemistry by Frank Pichardo (dec.) began his study and interest in science. Indeed it was the quantum mechanical basis of chemistry that eventually led him to study physics in college. He received a Bachelor of Science in physics and chemistry from Florida State University in 2003, where an opportunity to attain a minor in physics led him to take a modern physics course where the first chapter introduced him to special relativity. This course turned to many, with a major in physics also added. His years at Florida State led him to work on two separate physics research projects including contributed to the designing and testing of the HB, HO, and HE calibration boxes for the compact muon solenoid (CMS) experiment at the Large Hadron Collider (LHC) located at CERN under the supervision of Professor Vasken Hagopian and Maurizio Bertoldi. During that  same time, he contributed to helium atom surface scattering experiment by interpreting the surface structure and dynamics of TiO$_2$ under Professor James Skofronick, Professor Sanford Safron, and Dr.\ Elshan Akhadov. During the spring of 2003, Andr\'es accepted an opportunity to attend graduate school at  Louisiana State University and wrote his thesis under the guidance of Professor Gabriela Gonz\'alez in partial completion of the Master of Science degree in physics. 

\end{document}